\documentclass{article}

\usepackage{amsmath}
\usepackage{amssymb}
\usepackage{mathtools}
\usepackage{dsfont}
\usepackage{upgreek}
\usepackage[utf8]{inputenc}
\usepackage[english]{babel}
\usepackage[T1]{fontenc}
\usepackage[colorlinks,pdfpagelabels,pdfstartview = Fit,bookmarksopen = false,bookmarksnumbered = true,linkcolor = black,plainpages = false,hypertexnames = false,citecolor = black,urlcolor = blue] {hyperref} 
\usepackage{siunitx}
\usepackage{caption}
\usepackage{subcaption}
\usepackage{braket}
\usepackage{wrapfig}
\usepackage{amsthm}
\usepackage{paralist}
\usepackage{pgfplots}
\usepackage{standalone}

\usepackage{mathdesign}
\usepackage{textcomp}
\usepackage{geometry}
\usepackage{fancyhdr}
\usepackage{graphicx}
\usepackage{xcolor}
\usepackage{xkeyval}
\usepackage{eso-pic}

\newtheorem*{Def}{Definition}
\newtheorem{Thm}{Theorem}[section]
\newtheorem{Prop}[Thm]{Proposition}
\newtheorem{Cor}[Thm]{Corollary}
\newtheorem{Obs}[Thm]{Observation}

\newcommand{\B}{\mathcal{B}(\mathcal{H})}
\newcommand{\Hi}{\mathcal{H}}
\newcommand{\U}[1]{\mathcal{U}(#1)}
\newcommand{\I}{\mathds{1}}
\newcommand{\Tr}[1]{\mathrm{Tr}\left[#1\right]}
\newcommand{\GL}[1]{\mathrm{GL}(#1)}
\newcommand{\DistTo}{\xrightarrow{\,\smash{\raisebox{-0.65ex}{\ensuremath{\scriptstyle\sim}}}\,}}
\newcommand{\Int}[3]{\int \limits_{#1} #2 \mu(\mathrm{d}#3)}

\def\compileimages{1}

\hypersetup{%
  pdftitle={Construction of Random Unitary Operations for Asymptotic Preparation of Werner States},
  pdfauthor={Jakob Stonner},
  pdfsubject={Bachelor Thesis},
  pdfview=FitH,
  pdfstartview=FitV
}

\title{Bachelor Thesis: Construction of Random Unitary Operations for Asymptotic Preparation of Werner States}
\author{David Jakob Stonner
\\Institut für Angewandte Physik, 
\\Technische Universität Darmstadt,
\\Hochschulstraße 4a,
\\ D-64289 Darmstadt, Germany}
\date{13.08.2019}

\begin{document}

\maketitle

\abstract{Werner states are defined as bipartite qudit states that remain unchanged under application of arbitrary unitary operators acting on both subsystems simultaneously. Their preparation is a crucial ingredient in entanglement distillation protocols. This thesis deals with an iterative approach to prepare Werner states asymptotically, using random unitary operations. The asymptotic dynamics of random unitary operations are linked to algebraic properties of the involved unitary operators applying results about general quantum operations. Then a family of random unitary operations is constructed, which prepares Werner states asymptotically for an arbitrary finite dimensional bipartite qudit system. Finally, this construction is applied in qubit and qudrit systems, where the convergence rate is optimized numerically.}


\tableofcontents
\newpage
\section{Introduction}
Werner states are physical states of a bipartite quantum system that are invariant under application of arbitrary unitary operators acting on both subsystems simultaneously. Initially, Werner came up with this family of states in 1989 as an example of states admitting a hidden variable theory without being classically correlated \cite{werner}. Further on, because of their interesting properties, Werner states became frequently used examples in quantum information theory. Additionally, practical applications involving the preparation of Werner states are known, for instance in quantum state purification \cite{entanglement_purification,purification,purification_ndim}. The problem that is being addressed in this thesis is how to prepare these states.
\\ \\
An obvious approach is obtained by the fact that all Werner states are of the form
\begin{equation} \label{eq:U_average}
\Int{\U{d}}{U \otimes U\ X\ U^\dagger \otimes U^\dagger}{U},
\end{equation}
where $X$ is some bipartite state and $\mu$ is the Haar measure\footnote{This integral can be thought of as the generalization of averaging over group elements in the case of infinite groups.} on the locally compact Hausdorff group $\U{d}$ of unitary operators in dimension $d$. So if one picks a unitary $U \in \U{d}$ at random, let it act on the initial state $X$ according to equation \eqref{eq:U_average} and repeats this process over and over, eventually the resulting state will converge to a Werner state. This procedure can be thought of as computing the integral \eqref{eq:U_average} using the Monte Carlo method. However, this approach is not very practical since one would have to prepare a large number of unitary operators to act on the system.
\\ \\
This thesis deals with the approach of constructing a single quantum operation $T$ that prepares Werner states asymptotically, i.e. such that $T^n X$ converges to a Werner state for any initial state $X$ and $n \to \infty$. If $T$ can be implemented using just a few unitary operators then the effort reduces dramatically when compared to the first approach. Additionally, the usage of quantum operations allows for estimates about the quality of convergence. The construction of random unitary operations for the asymptotic preparation of Werner states will be presented as a practical application of the very general theory of quantum operations and their asymptotics. 
\\ \\
This work is organized as follows. In the beginning, chapter \ref{ch:basics} introduces all required basic concepts, including some theory about Werner states, a summary of results about quantum operations and their application in the special case of random unitary operations, as well as a brief discussion of elementary representation theory of finite groups. In chapter \ref{ch:werner_2d} we will apply the established foundations in order to construct random unitary operations that prepare Werner states in qubit systems. The next step is a generalization of the construction to arbitrary finite dimensional bipartite systems, which is essentially the subject of this thesis and will be addressed in chapter \ref{ch:werner_dd}. Finally, in chapter \ref{ch:application}, we will consider numerical examples in bipartite qubit and qudrit systems and discuss some convergence issues.
\\ \\
All numerical evaluations were done using Python 3.5.2, in particular the packages NumPy and SciPy. For all numerical computations a precision of $10^{-8}$ was chosen arbitrarily. All plots were created using the \LaTeX\ package PGFPLOTS. For group theoretic analysis the computer algebra system GAP 4.10.1 was used. This thesis was handed in on 13.08.2019 and has been edited afterwards.
\section{Basic Concepts}\label{ch:basics}
In the following we build up the required concepts that we are going to utilize for the asymptotic preparation of Werner states. We start in section \ref{sec:werner} with defining Werner states properly and collect some of their properties, in particular introducing the Haar measure for averaging over infinite groups in section \ref{sec:twirling}. After that follows the introduction of quantum operations in section \ref{sec:quantum_op} and some results about their asymptotic behaviour are presented. We will then apply those in chapter \ref{sec:ruo} in the special case of random unitary operations. At last, in section \ref{sec:representations} the basic notions of representation theory for finite groups are being introduced, which will in the end allow for an elegant formulation and proof of a structure result on asymptotic eigenvectors.
\subsection{Werner States} \label{sec:werner}
We consider a bipartite quantum system of two $d$ dimensional Hilbert spaces $\Hi = \Hi_\mathrm{A} \otimes \Hi_\mathrm{B}$. This describes a system of two particles which can be in $d$ distinct states individually. We denote the space of bounded operators on $\Hi$ as $\B$ and the space of unitary operators in dimension $d$ as $\U{d}$.
In this setting we define Werner states as follows:
\begin{Def}
A state $\rho \in \B$ is called \textbf{Werner state} if
\begin{equation*}
U \otimes U\ \rho\ U^\dagger \otimes U^\dagger = \rho \text{ for all }U \in \mathcal{U}(d).
\end{equation*}
\end{Def}
Since operators of the form $U \otimes U$ will be used quite frequently throughout this work, we introduce the shorthand notation
\begin{equation*}
U_\otimes \coloneqq U \otimes U
\end{equation*}
for $U \in \U{d}$. Note that $(U^\dagger)_\otimes = (U_\otimes)^\dagger \eqqcolon U_\otimes^\dagger$ and $(U_\otimes)(V_\otimes) = (UV)_\otimes$. We write $U_\otimes.U_\otimes^\dagger$ for the map
\begin{equation*}
\B \to \B,\ X \mapsto U_\otimes X U_\otimes^\dagger.
\end{equation*}
Using this notation, Werner states are precisely those states that are invariant under $U_\otimes.U_\otimes^\dagger$ for all $U \in \U{d}$. In section \ref{sec:twirling_werner} we will see that equation \eqref{eq:U_average} gives the general form of a Werner state. However, there is a more useful expression for Werner states. First define the flip operator
\begin{equation}\label{eq:flip}
F: \Hi \to \Hi,\ \ket{\phi} \otimes \ket{\psi} \mapsto \ket{\psi} \otimes \ket{\phi},
\end{equation}
which flips the particle states of A and B. Now define the symmetric and asymmetric projection operator as
\begin{align} \label{eq:projection_ops}
\begin{split}
P_\mathrm{sym} &\coloneqq \frac{1}{2}( \I + F ) \\
P_\mathrm{asym} &\coloneqq \frac{1}{2}( \I - F ),
\end{split}
\end{align}
where $\I$ denotes the identity. The operator $P_\mathrm{sym}$ projects onto the eigenspace $\ker( F - \I )$, whereas $P_\mathrm{asym}$ projects onto the eigenspace $\ker( F + \I )$. We find that any Werner state is a linear combination of these projections.
\begin{Thm}\label{thm:werner}
For every Werner state $\rho \in \B$ there exists some $p \in [0,1]$ such that
\begin{equation}\label{eq:werner}
\rho = p \frac{2}{d(d+1)} P_\mathrm{sym} + (1-p)\frac{2}{d(d-1)} P_\mathrm{asym}.
\end{equation}
\end{Thm}
We prove this result using a variation of the original proof given by Werner\cite{werner}. Fix some orthonormal basis $\ket{1},...,\ket{d}$ of $\Hi_\mathrm{A}$ or $\Hi_\mathrm{B}$ respectively. We define an orthonormal basis of $\Hi$ by
\begin{align} \label{eq:werner_basis}
\begin{split}
\ket{\phi_i} &\coloneqq \ket{i}\ket{i}, \\
\ket{\phi_{i,j}} &\coloneqq \frac{1}{\sqrt{2}}( \ket{i}\ket{j} + \ket{j}\ket{i} ) \\
\ket{\psi_{i,j}} &\coloneqq \frac{1}{\sqrt{2}}( \ket{i}\ket{j} - \ket{j}\ket{i} ),
\end{split}
\end{align}
where $i,j \in \{1,...,d\}$ and $i<j$. This defines an eigenbasis of $F$, where $\ket{\phi_i}$ and $\ket{\phi_{ij}}$ have eigenvalue 1 while $\ket{\psi_{ij}}$ has eigenvalue -1. It follows that $P_\mathrm{sym}$ and $P_\mathrm{asym}$ are given by
\begin{align}\label{eq:projections_d}
\begin{split}
P_\mathrm{sym} &= \sum \limits_i \ket{\phi_i}\bra{\phi_i} + \sum \limits_{i<j} \ket{\phi_{i,j}}\bra{\phi_{i,j}} \\
P_\mathrm{asym} &= \sum \limits_{i<j} \ket{\psi_{i,j}}\bra{\psi_{i,j}}.
\end{split}
\end{align}
In particular, we find 
\begin{align} \label{eq:projection_trace}
\begin{split}
\dim \ker(F+\I) &= \Tr{P_\mathrm{asym}} = \sum \limits_{n=1}^{d-1} n = \frac{d(d-1)}{2} \\
\dim \ker(F-\I) &= \Tr{P_\mathrm{sym}} = d + \frac{d(d-1)}{2} = \frac{d(d+1)}{2}.
\end{split}
\end{align}
Now we prove Theorem \ref{thm:werner}.
\begin{proof}
To begin with, note that $P_\mathrm{sym}$ and $P_\mathrm{asym}$ commute with any $U_\otimes$ for $U \in \U{d}$ as $F$ does, which can easily be checked explicitly using some orthonormal basis. This shows that equation \eqref{eq:werner} indeed defines a Werner state for any $p \in [0,1]$.
\\ \\
Let $\rho \in \B$ be a Werner state. It suffices to show that $\rho$ is diagonal w.r.t. the basis \eqref{eq:werner_basis} and that the diagonal matrix elements corresponding to $\ket{\phi_i}$ and $\ket{\phi_{i,j}}$ as well as those corresponding to $\ket{\psi_{i,j}}$ coincide. For diagonality, observe that for any $i \in \{1,...,d\}$ there exists some unitary operator $U_i \in \U{d}$ sending $\ket{i}$ to $-\ket{i}$ while leaving all other basis vectors unchanged. For $i \neq j$ it follows
\begin{equation*}
\bra{\phi_i} \rho \ket{\phi_j} = \bra{\phi_i} (U_i)_\otimes^\dagger \rho (U_i)_\otimes \ket{\phi_j} = - \bra{\phi_i} \rho \ket{\phi_j} = 0
\end{equation*}
and analogously, for all non-diagonal matrix elements there exists a suitable $U_i$, showing that $\rho$ is indeed diagonal. Next we show $\bra{\phi_i} \rho \ket{\phi_i} = \bra{\phi_1} \rho \ket{\phi_1}$ and $\bra{\phi_{i,j}} \rho \ket{\phi_{i,j}} = \bra{\phi_{1,2}} \rho \ket{\phi_{1,2}}$. For any $i<j$ there exists a unitary operator $U_{ij}$ which exchanges $\ket{i}$ and $\ket{j}$ while leaving all remaining basis vectors unchanged. Thus for any $i \neq 1$ we have
\begin{equation*}
\bra{\phi_i} \rho \ket{\phi_i} = \bra{\phi_i} (U_{1i})_\otimes^\dagger \rho (U_{1i})_\otimes \ket{\phi_i} = \bra{\phi_1} \rho \ket{\phi_1},
\end{equation*}
and the same argument applies to the diagonal matrix elements corresponding to $\ket{\phi_{i,j}}$ and $\ket{\psi_{i,j}}$ if $2 \neq i<j$, using a unitary operator that exchanges $\ket{i}$ and $\ket{1}$ as well as $\ket{j}$ and $\ket{2}$ simultaneously. If $i = 2 <j$ then one can use the unitary implementation of the three cycle $(21j)$, i.e. $\ket{2} \mapsto \ket{1} \mapsto \ket{j} \mapsto \ket{2}$.
\\ \\
It remains to show $\bra{\phi_1} \rho \ket{\phi_1} = \bra{\phi_{1,2}} \rho \ket{\phi_{1,2}}$. Consider $U_\varphi = R_\varphi \oplus \I_{d-2}$ where $R_\varphi$ is a two dimensional rotation on $\mathrm{span}(\ket{1},\ket{2})$ given by its matrix representation
\begin{equation*}
R_\varphi = \begin{pmatrix}
\cos\varphi & -\sin\varphi \\
\sin\varphi & \cos\varphi \\
\end{pmatrix}.
\end{equation*}
If we choose $\varphi$ such that $\sin \varphi \neq 0 \neq \cos \varphi$ then it follows
\begin{align*}
\bra{\phi_1} \rho \ket{\phi_1} &= \bra{\phi_1} (R_\varphi)_\otimes^\dagger \rho (R_\varphi)_\otimes \ket{\phi_1} \\
&= \cos^4 \varphi\bra{\phi_1} \rho \ket{\phi_1} + \sin^4 \varphi\underbrace{\bra{\phi_2} \rho \ket{\phi_2}}_{=\bra{\phi_1} \rho \ket{\phi_1}} + 2 \sin^2\varphi\cos^2\varphi \bra{\phi_{1,2}} \rho \ket{\phi_{1,2}} \\
&= \underbrace{\frac{2 \cos^2 \varphi \sin^2 \varphi}{1-\sin^4 \varphi - \cos^4 \varphi}}_{=1} \bra{\phi_{1,2}} \rho \ket{\phi_{1,2}}.
\end{align*}
This finally shows $\rho \in \mathrm{span}(P_\mathrm{sym}, P_\mathrm{asym})$. Since $\rho$ is a physical state we require $\Tr{\rho} = 1$, therefore by equation \eqref{eq:projection_trace} the desired form follows.
\end{proof}
Note that we used a large number of unitary operators during the proof, depending in particular on the dimension $d$. Later we will see that surprisingly only three unitary operators suffice to obtain this result in arbitrary dimensions\footnote{In fact, we will later prove that if $d$ is odd then only two operators suffice, which might even be true without the restriction on $d$.}.
\subsection{Haar Measure and Twirling}\label{sec:twirling}
Finite groups possess the notion of averaging over all group elements. If $f: G \to \mathbb{C}^n$ is a map on the finite group $G$ then the average of $f$ under $G$ is implemented as
\begin{equation*}
\mathfrak{a}_G(f) \coloneqq \frac{1}{|G|} \sum \limits_{g \in G} f(g) \in \mathbb{C}^n.
\end{equation*}
\begin{Obs}\label{obs:group_average_props}
The following properties of $\mathfrak{a}_G(f)$ hold:
\begin{itemize}
\item{The map $f \mapsto \mathfrak{a}_G(f)$ is linear.}
\item{For all $g \in G$ and any translation\footnote{For finite groups this is even true for all bijective $\tau:G \to G$.} $\tau_g: h \mapsto gh$ we have $\mathfrak{a}_G(f \circ \tau_g) = \mathfrak{a}_G(f)$.}
\item{If $f \equiv c \in \mathbb{C}^n$ is constant then $\mathfrak{a}_G(f) = c$.}
\end{itemize}
\end{Obs}
This concept can be generalized to arbitrary locally compact Hausdorff groups. This is done by constructing a measure on such a group that has certain invariance properties, and thus gives rise to an integral that has similar properties like $\mathfrak{a}_G$. The following presentation is based on \cite{measuretheory}. In order to be precise, we recall some basic definitions from topology and topological group theory:
\begin{Def}
\begin{compactenum}[(a)]
\item{A topological space is called \textbf{Hausdorff} if all distinct points can be separated by disjoint open sets.}
\item{A topological space is called \textbf{locally compact} if every point possesses a compact neighbourhood.}
\item{A group $G$ together with some topology is called \textbf{topological group} if the inversion map $g \mapsto g^{-1}$ is continuous. We denote with $\mathfrak{B}$ the set of Borel sets on $G$, which is defined to be the $\sigma$-field generated by all open sets.}
\end{compactenum}
\end{Def}
We will only consider cases where $G$ is a subgroup of $\U{n}$, so it is important to note that indeed $\U{n}$ is a locally compact Hausdorff group\footnote{In fact, $\U{n}$ is even compact.}, and therefore so is every subgroup.
\begin{Thm}
Let $G$ be a locally compact Hausdorff group. Then there exists a unique measure $\mu: \mathfrak{B} \to \mathbb{R}$ satisfying the properties:
\begin{itemize}
\item{$\mu$ is left invariant, i.e. $\mu(gE) = \mu(E)$ for all $E \in \mathfrak{B}, g \in G$.}
\item{$\mu(G) = 1$}
\item{$\mu$ is quasiregular, i.e. inner regular for all measurable sets and outer regular for all \emph{open} measurable sets.}
\end{itemize}
The measure $\mu$ is called the \textbf{Haar measure} on $G$.
\end{Thm}
Now we set for any measurable map $f: G \to \mathbb{C}^n$, in analogy to the finite case:
\begin{equation*}
\mathfrak{a}_G(f) \coloneqq \int \limits_G f(g) \mu(\mathrm{d}g)
\end{equation*}
We find, that all properties stated in Observation \ref{obs:group_average_props} carry over.
\subsubsection{Twirling and Werner States}\label{sec:twirling_werner}
\begin{Def}
Let $\Hi$ be an $n$ dimensional Hilbert space, $G$ be a subgroup of $\U{n}$. We define the \textbf{twirling operation} of $G$ on $\B$ as
\begin{equation}\label{eq:twirling}
\mathfrak{t}_G:\B \to \B,\ X \mapsto \Int{G}{U X U^\dagger}{U},
\end{equation}
where $\mu$ is the Haar measure on $G$. The images of $\mathfrak{t}_G$ are called \textbf{twirling states} of $G$.
\end{Def}
Note that $\B \simeq \mathbb{C}^{(n,n)} \simeq \mathbb{C}^{n^2}$, $\mathcal{B}(\B) \simeq \mathbb{C}^{n^4}$ and $\mathfrak{t}_G = \mathfrak{a}_G(f)$ for
\begin{equation*}
f:G \to \mathcal{B}(\B),\ U \mapsto U.U^\dagger.
\end{equation*}
We therefore obtain the following property of twirling operations.
\begin{Obs}\label{obs:twirling}
Let $\mathfrak{t}_G$ be the twirling operation of $G \subseteq \U{n}$ on $\B$, let $U \in G$. Then we have
\begin{equation*}
U \mathfrak{t}_G(X) U^\dagger = \mathfrak{t}_G(X) \text{ for all }X \in \B.
\end{equation*}
\end{Obs}
\begin{proof}
Let $\mu$ be the Haar measure of $G$, let $\tau_U$ be the right translation of $U$. We find for all $X \in \B$:
\begin{equation*}
U \mathfrak{t}_G(X) U^\dagger = \Int{G}{UV X \underbrace{V^\dagger U^\dagger}_{=(UV)^\dagger}}{V} = \mathfrak{a}_G(f \circ \tau_U)(X) =  \mathfrak{a}_G(f)(X) = \mathfrak{t}_G(X)
\end{equation*}
\end{proof}
We can now prove rigorously that equation \eqref{eq:U_average} indeed defines Werner states in general.
\begin{Prop}\label{obs:werner_projection}
As in section \ref{sec:werner}, let $\Hi = \Hi_\mathrm{A} \otimes \Hi_\mathrm{B} \simeq \mathbb{C}^d \otimes \mathbb{C}^d$ and set for convenience of notation
\begin{equation*}
\mathcal{U} \coloneqq \Set{U_\otimes|U \in \U{d}} \simeq \U{d}.
\end{equation*}
Then Werner states are precisely the states $\mathfrak{t}_\mathcal{U}(X)$ for $X \in \B$.
\end{Prop}
\begin{proof}
If $\rho \in \B$ is a Werner state then 
\begin{equation*}
\mathfrak{t}_\mathcal{U}(\rho) = \Int{\mathcal{U}}{U_\otimes \rho U_\otimes^\dagger}{U_\otimes} = \Int{\mathcal{U}}{\rho}{U_\otimes} = \rho \mu(\mathcal{U}) = \rho.
\end{equation*}
Conversely, for all $X \in \B$ and $U_\otimes \in \mathcal{U}$ we find $U_\otimes \mathfrak{t}_\mathcal{U}(X) U_\otimes^\dagger = \mathfrak{t}_\mathcal{U}(X)$ by Observation \ref{obs:twirling}.
\end{proof}
\subsection{Quantum Operations}\label{sec:quantum_op}
Let $\Hi$ be an finite dimensional Hilbert space over $\mathbb{C}$ with scalar product $\braket{.|.}$, let $\B$ be the Hilbert space of bounded operators on $\Hi$ with the Hilbert Schmidt scalar product $\braket{A,B}_\mathrm{HS} = \Tr{A^\dagger B}$. If not stated otherwise, throughout this section $\B$ will be equipped with the norm that is induced by the Hilbert Schmidt scalar product. As $\Hi$ is isomorphic to the unitary space $\mathbb{C}^n$ for $n = \dim( \Hi )$, $\B$ is isomorphic to the space $\mathbb{C}^{(n,n)}$ of $n \times n$ matrices.
\\ \\
The notion of quantum operations is that of mapping states to states within $\B$. As states are defined to be the matrices $X \in \B$ with $X \geq 0$ and $\Tr{X} = 1$, such an operator should preserve positivity and be trace non-increasing. 
\begin{Def}
A linear map $T: \B \to \B$ is called \textbf{quantum operation} if it is trace non-increasing and completely positive, i.e. 
for all $k \in \mathbb{N}_0$ and $X \in \mathbb{C}^{(k,k)} \otimes \B$ it holds
\begin{equation*}
X \geq 0\ \Rightarrow\ ( \I_{(k,k)} \otimes T )(X) \geq 0.
\end{equation*}
If $T$ is additionally trace preserving, it is called \textbf{quantum channel}.
\end{Def}
The following proposition gives a useful characterization of these properties. A proof can be found in \cite{quantuminformationscience}.
\begin{Prop}
Let $T: \B \to \B$ be an linear map.
\begin{compactenum}[(a)]
\item{$T$ is trace non-increasing if and only if $T(\I) \leq \I$. $T$ is trace preserving if and only if $T(\I) = \I$. }
\item{$T$ is completely positive if and only if there exist operators $A_1,...,A_k \in \B$ such that
\begin{equation*}
T(X) = \sum \limits_{i=1}^k A_i X A^\dagger_i \text{ for all } X \in \B. 
\end{equation*}
The $A_i$ are called \textbf{Kraus operators} of $T$. }
\end{compactenum}
\end{Prop}
\begin{Obs}
Let $T$ be an quantum operation.
\begin{compactenum}[(a)]
\item{The spectrum $\sigma$ of $T$ is bounded by 1, i.e. $|\lambda| \leq 1$ for all $\lambda \in \sigma$.}
\item{If $A_1,...,A_k$ are the Kraus operators of $T$ then the adjoint quantum operation $T^\dagger$ is given by
\begin{equation*}
T^\dagger(X) = \sum \limits_{i=1}^k A_i^\dagger X A_i \text{ for all } X \in \B. 
\end{equation*}
}
\end{compactenum}
\end{Obs}
What happens if we iterate such a map, apply it on an initial state and consider the limit? If the quantum operation is diagonalizable then the answer is obvious: All eigenvectors corresponding to eigenvalues with unit modulus determine the limit, all other eigenvectors vanish asymptotically. As it turns out, this remains true in the general case.
\subsubsection{Asymptotic Dynamics of Quantum Operations}\label{sec:quantum_op_asymp}
Let $T$ be a quantum operation, let $\sigma \subseteq \mathbb{C}$ be the set of all its eigenvalues and $\sigma_1 \coloneqq \Set{ \lambda \in \sigma | {|\lambda|} = 1 } \subseteq \sigma$ be the set of unit modulus eigenvalues. For any $\lambda \in \sigma$ let $d_\lambda \coloneqq \dim \ker( T - \lambda \I )$ be the geometric multiplicity of $\lambda$. We denote with $X_{\lambda,1},...,X_{\lambda,d_\lambda} \in \ker(T - \lambda \I)$ the $d_\lambda$ pairwise linearly independent eigenvectors and choose them to have unit norm. For $i \in \{ 1,...,d_\lambda\}$ we associate with $X_{\lambda,i}$ its dual vector $X^{\lambda,i} \in \B$ defined by
\begin{equation}
\braket{X^{\lambda,i},X_{\lambda',i'}}_\mathrm{HS} = \delta_{\lambda \lambda'} \delta_{i i'} \text{ for all }\lambda' \in \sigma,\ i' \in \{1,...,d_{\lambda'}\}.
\end{equation}
That such dual basis vectors always exist follows from Riesz theorem. We now obtain the asymptotic behaviour of $T$ explicitly as follows:
\begin{Thm}
Let $X_0 \in \B$ be some initial state. Define for $n \in \mathbb{N}$
\begin{equation}\label{eq:quantum_op_asym}
X_\infty(n) \coloneqq \sum \limits_{\lambda \in \sigma_1} \sum \limits_{i=1}^{d_\lambda} \lambda^n X_{\lambda,i} \braket{X^{\lambda,i}, X_0}_\mathrm{HS}.
\end{equation}
Then it holds $\| T^n(X_0) - X_\infty(n) \| \to 0$ for $n \to \infty$, where $\|.\|$ denotes the Hilbert Schmidt norm. More specifically, if we set
\begin{equation}\label{eq:maxeig}
\lambda_\mathrm{max} \coloneqq \mathrm{max}\{ |\lambda|\ |\ \lambda \in \sigma \setminus \sigma_1 \}
\end{equation}
then we obtain the following estimates:
\begin{compactenum}[(a)]
\item{If $T$ is diagonalizable then we have 
\begin{equation}\label{eq:estimate_diag}
\| T^n(X_0) - X_\infty(n) \| \leq C \lambda_\mathrm{max}^n
\end{equation}
for all $n \in \mathbb{N}$, where $C > 0$ is some positive constant.}
\item{If $T$ is not diagonalizable then we have 
\begin{equation}\label{eq:estimate_jordan}
\| T^n(X_0) - X_\infty(n) \| \leq C n^m \lambda_\mathrm{max}^{n-m}
\end{equation}
for all $n \geq m$, where $m$ is the greatest multiplicity of all eigenvalues in the minimal polynomial of $T$ and $C > 0$ is some positive constant.}
\end{compactenum}
\end{Thm}
As an immediate consequence of the theorem we obtain the following corollary:
\begin{Cor}\label{cor:asymp_stat}
For a quantum operation $T$ and $X_0 \in \B$ the following assertions are equivalent:
\begin{compactenum}[(a)]
\item{$T^n(X_0)$ converges for $n \to \infty$}
\item{$X_\infty(n)$ converges for $n \to \infty$}
\item{$\sigma_1 \subseteq \{1\}$}
\end{compactenum}
\end{Cor}
\begin{Def}
In the case that one of the equivalent conditions in Corollary \ref{cor:asymp_stat} is met, we call the quantum operation \textbf{asymptotically stationary}. Eigenvectors corresponding to eigenvalues of unit modulus will be called \textbf{asymptotic eigenvectors} subsequently. 
\end{Def}
The proof of the theorem for general quantum operations uses a matrix representation of $T$ in Jordan normal form, which always exists since its characteristic polynomial splits into linear factors over $\mathbb{C}$. The main result $\| T^n(X_0) - X_\infty(n) \| \to 0$ is proven in \cite{quantumops}. Since later discussions of convergence issues will involve the estimates \eqref{eq:estimate_diag} and \eqref{eq:estimate_jordan} which are not given in the paper, we carry out parts of the proof explicitly.
\begin{proof}
First consider the case where $T$ is diagonalizable. Then we can write
\begin{equation*}
T = \sum \limits_{\lambda \in \sigma} \lambda \underbrace{\sum \limits_{i=1}^{d_\lambda} X_{\lambda,i} \braket{X^{\lambda,i},.}_\mathrm{HS}}_{\eqqcolon P_\lambda} = \sum \limits_{\lambda \in \sigma} \lambda P_\lambda
\end{equation*}
where $P_\lambda$ is the (not necessarily orthogonal) projection onto the eigenspace $\ker(T-\lambda \I)$. It follows
\begin{align*}
\| T^n(X_0) - X_\infty(n) \| &= \| \sum \limits_{|\lambda| < 1} \lambda^n P_\lambda(X_0) \| \\
&\leq \sum \limits_{|\lambda| < 1} |\lambda|^n \| P_\lambda(X_0) \| \leq C \lambda_\mathrm{max}^n
\end{align*}
for some constant $C > 0$. Now let $T$ be an arbitrary quantum operation. We put $T$ in Jordan normal form and write
\begin{equation*}
T = \bigoplus \limits_{\lambda \in \sigma} \bigoplus \limits_{i=1}^{d_\lambda} J_{\lambda,i},
\end{equation*}
where $J_{\lambda,i}$ is a Jordan block of the form
\begin{equation*}
J_{\lambda,i} = \begin{pmatrix}
\lambda & 1 & 0 & \cdots & 0 \\
0 & \lambda & 1 & & \vdots \\
\vdots & & \ddots & \ddots & \\
 & & & \lambda & 1 \\
0 & \cdots & & 0 & \lambda
\end{pmatrix} \in \mathbb{C}^{(l_{\lambda,i}, l_{\lambda,i})}
\end{equation*}
in its corresponding basis and $ 1 \leq l_{\lambda,i} \leq m_\lambda$ with the multiplicity $m_\lambda$ of $\lambda$ in the minimal polynomial 
\begin{equation*}
q = \prod \limits_{\lambda \in \sigma} (X - \lambda)^{m_\lambda} \in \mathbb{C}[X].
\end{equation*}
Using induction, one can easily show 
\begin{equation*}
(J_{\lambda,i}^n)_{kl} = \binom{n}{l-k} \lambda^{n-(l-k)}
\end{equation*}
for all $n \in \mathbb{N}$, where we set $\binom{n}{k} \coloneqq 0$ if $k<0$ or $k>n$. Since $l-k \leq l_{\lambda,i} \leq m_\lambda$ this can be further estimated to
\begin{equation*}
|(J_{\lambda,i}^n)_{kl}| = \binom{n}{l-k} |\lambda|^{n-(l-k)} \leq n^{m_\lambda} |\lambda|^{n-m_\lambda},
\end{equation*}
which holds for all $n \geq m_\lambda$. Let $J_{\lambda,i} \in \mathbb{C}^{(l_{\lambda,i}, l_{\lambda,i})}$ for $\lambda \in \sigma$ and $i \in \{1,...,d_\lambda\}$ be a Jordan block and $X \in \mathbb{C}^{l_{\lambda,i}}$ be some vector. For all $n\geq m_\lambda$ we obtain an estimate in the supremum norm $\|.\|_\infty$ w.r.t. the Jordan basis of $T$ as
\begin{align*}
\|J_{\lambda,i}^n X\|_\infty &= \max \limits_{k} |(J_{\lambda,i}^n X)_k| = \max \limits_k |\sum \limits_{l} (J_{\lambda,i}^n)_{kl} X_l| \\
&\leq n^{m_\lambda} |\lambda|^{n-m_\lambda} \sum \limits_l | X_l| \\
&\leq n^{m_\lambda} |\lambda|^{n-m_\lambda} m_\lambda \|X\|_\infty.
\end{align*}
Now we use the fact proven in \cite{quantumops}, that all Jordan blocks corresponding to eigenvalues of unit modulus have dimension 1, i.e. $l_{\lambda,i} = 1$ for all $\lambda \in \sigma_1, i \in \{1,...,d_\lambda\}$. If we denote with $P_{\lambda,i}$ the projection onto the subspace of the eigenspace $\ker(T-\lambda \I)$ corresponding to $J_{\lambda,i}$, it follows
\begin{align*}
\| T^n(X_0) - X_\infty(n) \|_\infty &= \| \sum \limits_{|\lambda| < 1} \sum \limits_{i=1}^{d_\lambda} J_{\lambda,i}^n P_{\lambda,i}(X_0) \|_\infty \\
&\leq \sum \limits_{|\lambda| < 1} \sum \limits_{i=1}^{d_\lambda} \| J_{\lambda,i}^n P_{\lambda,i}(X_0) \|_\infty \\
&\leq C' n^m \lambda_\mathrm{max}^{n-m}
\end{align*}
for some constant $C' > 0$ and $m = \max_\lambda m_\lambda$. Now note that in finite dimensional vector spaces all norms are equivalent, which completes the proof.
\end{proof}
Note that if $T$ even is unitarily diagonalizable, i.e. there exists an orthonormal eigenbasis of $T$, then we can calculate a bound for the constant $C$ in equation \eqref{eq:estimate_diag} explicitly. In this case $P_\lambda$ is an orthogonal projection, therefore it has operator norm 1 and we find
\begin{equation}\label{eq:estimate_diag_onb}
\| T^n(X_0) - X_\infty(n) \| \leq \sum \limits_{|\lambda| < 1} |\lambda|^n \| P_\lambda(X_0) \| \leq \nu \| X_0 \| \lambda_\mathrm{max}^n
\end{equation}
with $\nu \coloneqq |\sigma \setminus \sigma_1|$. For a non-orthogonal projection $P_\lambda$ we can estimate
\begin{equation}\label{eq:eigenprojection}
\| P_\lambda X_0 \| \leq \sum \limits_{i=1}^{d_\lambda} |\braket{X^{\lambda,i}, X_0}_\mathrm{HS}| \cdot \| X_{\lambda,i} \| \leq \| X_0 \| \sum \limits_{i=1}^{d_\lambda} \| X^{\lambda,i} \|
\end{equation}
using the Cauchy-Schwarz inequality, thus if the dual basis vectors are known then a bound for $C$ can be calculated as well.
\\ \\
To determine the asymptotic dynamics it remains to calculate the dual eigenvectors, which can in general be a quite tedious problem since the eigenvectors of $T$ are not orthogonal w.r.t. each other in general. Fortunately, if $T$ possesses a positive state $0 < \rho \in \B$ with $T(\rho) \leq \rho$ one can find a dramatic simplification of the problem. The idea is to consider the inner product
\begin{equation*}
\braket{A, B}_\rho \coloneqq \braket{A,B \rho^{-1}}_\mathrm{HS} = \Tr{A^\dagger B \rho^{-1}}
\end{equation*}
instead of the Hilbert Schmidt scalar product. As it turns out, the eigenspaces of two distinct eigenvalues in $\sigma_1$ are orthogonal w.r.t. this new scalar product, which gives rise to an explicit expression for the dual eigenvectors.\cite{quantumops}
\begin{Thm}
Let $0 < \rho \in \B$ with $T(\rho) \leq \rho$, $\lambda \in \sigma_1$ and $i \in \{1,...,d_\lambda \}$. Then the dual eigenvector of $X_{\lambda,i}$ is given by
\begin{equation}\label{eq:dual_eigenvector}
X^{\lambda,i} = X_{\lambda,i} \rho^{-1} \braket{X_{\lambda,i},X_{\lambda,i}}_\rho^{-1}.
\end{equation} 
\end{Thm}
Now that we have a nice tool for dual eigenvectors, it remains to determine the actual asymptotic eigenvectors. If $\rho$ even is a fixed point of $T$ then one can show the following handy characterization of such eigenvectors, which links the problem to the individual Kraus operators.\cite{quantumops}
\begin{Thm}\label{thm:quantumop_eigenspace}
Let $A_1,...,A_k \in \B$ be the Kraus operators of $T$, let $0 < \rho \in \B$ be a fixed point of $T$ and let $\lambda \in \sigma_1$. Then $X \in \ker( T - \lambda \I )$ if and only if all of the following equations are satisfied for all $i \in \{1,...,k\}$:
\begin{align}\label{eq:attractor_equations}
\begin{split}
A_i X \rho^{-1} &= \lambda X \rho^{-1} A_i \\
A_i \rho^{-1} X &= \lambda \rho^{-1} X A_i \\
A_i^\dagger X \rho^{-1} &= \overline{\lambda} X \rho^{-1} A_i^\dagger \\
A_i^\dagger \rho^{-1} X &= \overline{\lambda} \rho^{-1} X A_i^\dagger
\end{split}
\end{align}
\end{Thm}
The equations \eqref{eq:attractor_equations} are known as \emph{attractor equations} and provide a powerful algebraic tool for determining the eigenspaces that contribute to asymptotic dynamics of a quantum operation.
\subsubsection{Random Unitary Operations}\label{sec:ruo}
For preparing Werner states in a bipartite System we are particularly interested in quantum operations with Kraus operators that are multiples of unitary operators. Such operations are called \emph{random unitary operations}.
\begin{Def}
Let $p_1,...,p_k \in (0,1]$ with $\sum_{i=1}^k p_i = 1$, let $U_1,...,U_k \in \U{n}$. Then the quantum operation
\begin{equation}\label{eq:random_unitary}
T: \B \to \B,\ X \mapsto \sum \limits_{i=1}^k p_i U_i X U_i^\dagger
\end{equation}
is called \textbf{random unitary operation}.
\end{Def}
We will refer to the $U_i$ as the \emph{unitary Kraus operators} of $T$, despite the fact that the Kraus operators of $T$ are actually $\sqrt{p_i} U_i$. There is a straightforward way of implementing random unitary operations practically, which also justifies the name: One prepares every unitary operator $U_j$ to act on the system with a probability $p_j$. After a sufficient amount of random applications the resulting state will then be close to the iterative application of $T$.
\\ \\
Note that any random unitary operation $T$ is a quantum channel, since $T(\I) = \sum_{i=1}^k p_i U_i U_i^\dagger = \I$. In particular $\I$ is a positive fixed point of $T$, and we can significantly simplify the results obtained in section \ref{sec:quantum_op_asymp}. To begin with, for $\rho = \I$ the inner product $\braket{.,.}_\rho$ coincides with the Hilbert Schmidt scalar product and therefore equation \eqref{eq:dual_eigenvector} implies $X^{\lambda,i} = X_{\lambda,i}$ for all $\lambda \in \sigma_1$, i.e. all distinct asymptotic eigenvectors are orthogonal to each other. Using this fact, Theorem \ref{thm:quantumop_eigenspace} with Kraus operators $A_i = \sqrt{p_i} U_i$ yields this corollary:
\begin{Cor}\label{cor:eigenspace_intersection}
Let $T$ be a random unitary operation as in equation \eqref{eq:random_unitary}, let $\lambda \in \sigma_1$. Then $X \in \ker( T - \lambda\I )$ if and only if we have $U_j X U_j^\dagger = \lambda X$ for all $j \in \{1,...,k\}$. Equivalently we can write
\begin{equation*}
\ker( T - \lambda \I ) = \bigcap\limits_{i=1}^k \ker( U_i . U_i^\dagger - \lambda \I ).
\end{equation*}
\end{Cor}
Note that in particular asymptotic eigenvectors are independent of the choice of coefficients $p_i$. Now we go a step further and investigate the eigenspace for the eigenvalue 1, since in the case of a stationary asymptotic behaviour ($\sigma_1 \subseteq \{1\}$) this is the only eigenspace we care about. What can we say about the eigenvectors? Surprisingly, we find a connection to the group that is generated by the unitary Kraus operators. Define $G \coloneqq \braket{U_1,...,U_k} \subseteq \U{n}$, which is an at most countable subgroup of the unitary group. Then we can show that the eigenvectors corresponding to the eigenvalue 1 are precisely the twirling states of $G$ on $\B$ (recall section \ref{sec:twirling_werner}).
\begin{Thm}\label{thm:eigenspace_group}
Let $T$ be a random unitary operation of the form \eqref{eq:random_unitary}. Then the eigenspace for eigenvalue 1 is given by
\begin{equation*}
\ker(T - \I ) = \bigcap\limits_{i=1}^k \ker( U_i . U_i^\dagger - \I ) = \Set{ \mathfrak{t}_G(X) | X \in \B},
\end{equation*}
where $\mathfrak{t}_G$ is the twirling operation of $G$ on $\B$, as defined in equation \eqref{eq:twirling}.
\end{Thm}
\begin{proof}
If $X \in \ker(T - \I )$ then we have $U_j X U_j^\dagger = X$ for all $j \in \{1,...,k\}$. Since the $U_j$ generate $G$, this implies $U X U^\dagger = X$ for all $U \in G$. It follows
\begin{equation*}
\mathfrak{t}_G(X) = \Int{G}{U\ X\ U^\dagger}{U} = \Int{G}{X}{U} = X \mu(G) = X.
\end{equation*}
Conversely, if $X \in \B$ then by Observation \ref{obs:twirling} we find for all $j \in \{1,...,k\}$
\begin{equation*}
U_j\ \mathfrak{t}_G(X) U_j^\dagger = \mathfrak{t}_G(X),
\end{equation*}
thus $\mathfrak{t}_G(X) \in \bigcap_{j=1}^k \ker( U_j . U_j^\dagger - \I ) = \ker(T - \I )$.
\end{proof}
In particular the group $G$ determines the eigenspace $\ker(T - \I )$, and if two random unitary operations have the same group associated with them, then they share this eigenspace as well.
\subsection{Representation Theory of Finite Groups}\label{sec:representations}
In abstract algebra, group theory provides a frame for a very general understanding of basic structure, that has proven itself to be very useful throughout mathematics. Due to its abstract nature, it is not obvious how to utilize its results in real world applications that usually take place in vector spaces. This is where representation theory comes in. In the following let $G$ be a finite\footnote{It is possible to extend most of the results presented here to infinite groups, which is not needed here.} group and $V$ be a complex vector space. We denote with $\GL{V}$ the automorphism group of $V$, i.e. the set of all bijective linear operators on $V$.




\begin{Def}
The pair $(V, \varphi)$ is called a (linear) \textbf{representation} of $G$ if $\varphi: G \to \GL{V}$ is a group homomorphism, i.e. $\varphi(g)\varphi(h) = \varphi(gh)$ for all $g,h \in G$. The dimension of $V$ is called the \textbf{degree} of the representation. Occasionally the representation $(V, \varphi)$ is referred to just by $\varphi$.
\end{Def}
In particular we have $\varphi(e) = \I$ and $\varphi(g^{-1}) = \varphi(g)^{-1}$ by the homomorphism property ($e$ denotes the neutral element of $G$).
\\ \\
For illustration purposes we consider the following examples of group representations:
\begin{compactenum}[(a)]
\item{The representation obtained by setting $\varphi \equiv \I$ is called the \emph{trivial representation}.}
\item{Let $V$ have a basis $(e_g)_{g \in G}$ indexed by the elements of $G$ (in particular $\dim( V ) = |G|$). Now define $\varphi : G \to \GL{V}$ by
\begin{equation*}
\varphi( g )(e_h) \coloneqq e_{gh} \text{ for all }g, h \in G.
\end{equation*}
Then the corresponding representation is called the \emph{regular representation of $G$} (note that any linear map is determined by its action on the basis vectors).}
\item{Let $G$ be generated by finitely many elements $g_1,...,g_n$ of a larger group $H$, let $V$ be some complex vector space. Now choose automorphisms $A_1,...,A_n \in \GL{V}$ and set $\varphi(g_i) \coloneqq A_i$ for $i \in \{1,...,n\}$. This defines a unique representation $(V,\varphi)$, as the homomorphism property of $\varphi$ requires
\begin{equation*}
\varphi( g_{i_1}^{n_1}...\ g_{i_k}^{n_k} ) = \varphi(g_{i_1})^{n_1}...\ \varphi(g_{i_k})^{n_k} \text{ for all } i_1,...,i_k \in \{1,...n\}, n_1,...,n_k \in \mathbb{Z}.
\end{equation*}}
\end{compactenum}
\begin{Def}
Two representations $(V, \varphi)$ and $(V', \varphi')$ of $G$ are said to be \textbf{equivalent} (or similar, isomorphic) if there exists a vector space isomorphism $\tau : V \DistTo V'$ with $\tau \circ \varphi(g) = \varphi'(g) \circ \tau$ for all $g \in G$.
\end{Def}
If $V$ is finite dimensional then images of $\varphi$ can be identified with regular matrices w.r.t. some fixed basis. In this case another way of saying that $\varphi$ and $\varphi'$ are equivalent is that there exists a matrix $T \in \mathrm{GL}_n( \mathbb{C} )$ such that $T \varphi(g) T^{-1} = \varphi'(g)$ for all $g \in G$. In particular, two one dimensional representations are equivalent if and only if they are identical.
\begin{Def}
Let $(V, \varphi)$ be a representation of $G$. 
\begin{compactenum}[(a)]
\item{Let $W \subseteq V$ be a linear subspace that is stable under the action of $G$, i.e. $\varphi(g)(w) \in W$ for all $g \in G, w \in W$. Then $(W, \varphi|_W)$ is a representation of $G$ as well, which is called the \textbf{subrepresentation} of $(V, \varphi)$ induced by $W$. Here $\varphi|_W$ is given by $\varphi|_W(g) \coloneqq \varphi(g)|_W \in \GL{W}$. The subrepresentation $\varphi|_W$ is called trivial if $W \in \{ \{0\}, V \}$.}
\item{Let $\varphi|_{W_1},...,\varphi|_{W_k}$ be subrepresentations of $\varphi$. Then $\varphi$ is said to be the direct sum of $\varphi|_{W_1},...,\varphi|_{W_k}$ if $V = \bigoplus_{i = 1}^k W_i$.}
\item{$\varphi$ is called \textbf{reducible} if there exists a decomposition of $\varphi$ into a direct sum of two or more non-trivial subrepresentations. Otherwise it is called \textbf{irreducible}.}
\end{compactenum}
\end{Def}
We consider some of the most important results from elementary representation theory. The proofs can be found in any book about representation theory, for instance \cite{representations}.
\begin{Thm}
Any representation of finite degree decomposes into a direct sum of irreducible representations.
\end{Thm}
\begin{Thm}[Schur's lemma]
Let $(V_1, \varphi_1)$ and $(V_2, \varphi_2)$ be irreducible representations of $G$, let $f:V_1 \to V_2$ be linear with
\begin{equation*}
\varphi_2(g) \circ f = f \circ \varphi_1(g) \text{ for all }g \in G.
\end{equation*}
\begin{compactenum}[(a)]
\item{If $\varphi_1$ and $\varphi_2$ are not isomorphic then $f \equiv 0$.}
\item{If $V_1 = V_2$ and $\varphi_1 = \varphi_2$ then $f = c \I$ for some $c \in \mathbb{C}$.}
\end{compactenum}
In particular all irreducible representations of abelian groups are one dimensional.
\end{Thm}
Now we have the tools ready to prove the following result, which will be a key observation for constructing random unitary operations with specific asymptotic eigenvectors.
\begin{Thm}\label{thm:eigenvector_diagonal}
Let $(V, \varphi)$ be a representation of $G$ of degree $n \in \mathbb{N}$ that decomposes into a direct sum of one dimensional subrepresentations. Let $B = (\ket{b_1},...,\ket{b_n})$ be the basis of $V$ formed by the normalized basis vectors that span the one dimensional subspaces corresponding to the irreducible subrepresentations. If the subrepresentations are pairwise inequivalent then 
\begin{equation*}
\overline{X} \coloneqq \frac{1}{|G|} \sum \limits_{g \in G} \varphi( g )\ X\ \varphi( g )^{-1}
\end{equation*}
is diagonal w.r.t. $B$ for all $X \in \mathcal{B}(V)$.
\end{Thm}
\begin{proof}
For any $i \in \{1,...,n\}$ we write the corresponding irreducible representation as 
\begin{equation*}
\varphi|_{\mathrm{span}(\ket{b_i})}( g )(\ket{b_i}) = \phi_i(g) \ket{b_i} \text{ for all }g \in G,
\end{equation*}
where $\phi_i : G \to \mathbb{C} \setminus \{0\}$ can be thought of as the $1 \times 1$ matrix representation of $\varphi|_{\mathrm{span}(\ket{b_i})}$. Now fix $i \in \{1,...,n\}$ and expand $\overline{X} \ket{b_i}$ in the basis $B$ as $\overline{X} \ket{b_i} = \sum_k \lambda_k \ket{b_k}$. For all $g \in G$ and $j \in \{1,...,n\}$ it holds
\begin{equation*}
\bra{b_j}\varphi( g )\overline{X} \ket{b_i} = \bra{b_j} \sum \limits_k \lambda_k \varphi( g ) \ket{b_i}  = \bra{b_j} \sum \limits_k \lambda_k \phi_k( g ) \ket{b_i} = \lambda_j \phi_j(g).
\end{equation*}
Suppose we have $\lambda_j \neq 0$ for some $j \in \{1,...,n\}$. We claim that this implies $\phi_i \equiv \phi_j$, which then yields $i = j$ since $\phi_i$ and $\phi_j$ are assumed to be inequivalent for $i \neq j$. We find for all $g \in G$
\begin{align*}
\phi_j(g) &= \lambda_j^{-1} \bra{b_j} \varphi( g )\overline{X} \ket{b_i} \\
&= \lambda_j^{-1} \bra{b_j} \frac{1}{|G|} \sum \limits_{h \in G} \varphi( g ) \varphi( h )\ X\ \varphi( h )^{-1} \ket{b_i} \\
&= \lambda_j^{-1} \bra{b_j} \frac{1}{|G|} \sum \limits_{h \in G} \varphi( g h )\ X\ \varphi( gh )^{-1} \varphi(g) \ket{b_i} \\
&= \lambda_j^{-1} \bra{b_j} \overline{X} \phi_i(g) \ket{b_i} \\
&= \lambda_j^{-1} \phi_i(g) \underbrace{\bra{b_j} \overline{X} \ket{b_i}}_{= \lambda_j} \\
&= \phi_i(g).
\end{align*}
\end{proof}
In particular all matrix elements corresponding to inequivalent one dimensional representations vanish. Note that if $V \subseteq \U{n}$ then $\varphi(G) \subseteq \U{n}$ is a finite subgroup and we have $\overline{X} = \mathfrak{t}_{\varphi(G)}(X)$, which draws a connection towards the asymptotic eigenvectors of random unitary operations (recall Theorem \ref{thm:eigenspace_group}).
\section{Preparation of Werner States in Bipartite Qubit Systems}\label{ch:werner_2d}
The following construction is based on calculations from Prof. Dr. Alber, which he provided through private communication. We consider a bipartite qubit system, represented by a four dimensional Hilbert space $\Hi = \Hi_\mathrm{A} \otimes \Hi_\mathrm{B}$ with $\dim( \Hi_\mathrm{A} ) = \dim( \Hi_\mathrm{B}) = 2$. We fix some orthonormal basis $(\ket{1}, \ket{2})$ of $\Hi_\mathrm{A}$ (or $\Hi_\mathrm{B}$ resp.) and work in the basis of $\Hi$ that is given by the Bell states
\begin{align*}
\begin{split}
\ket{\Phi_\pm} &\coloneqq \frac{1}{\sqrt{2}}( \ket{1}\ket{1} \pm \ket{2}\ket{2}) \\
\ket{\Psi_\pm} &\coloneqq \frac{1}{\sqrt{2}}( \ket{1}\ket{2} \pm \ket{2}\ket{1}).
\end{split}
\end{align*}
The projection operators defined in equation \eqref{eq:projection_ops} are given by $P_\mathrm{sym} = \ket{\Phi_+}\bra{\Phi_+} + \ket{\Phi_-}\bra{\Phi_-} + \ket{\Psi_+}\bra{\Psi_+}$ and $P_\mathrm{asym} = \ket{\Psi_-}\bra{\Psi_-}$, thus Werner states by Theorem \ref{thm:werner} take the form
\begin{equation}\label{eq:werner_2d}
\rho = \frac{p}{3} P_\mathrm{sym} + (1-p) P_\mathrm{asym}
\end{equation}
for $p \in [0,1]$. Note that Werner states are therefore diagonal in the Bell basis. Now we consider the unitary operators
\begin{align}
\begin{split}
h^{(1)} &\coloneqq i \boldsymbol\sigma_3 = \begin{pmatrix} i & 0 \\ 0 & -i \\ \end{pmatrix} \\
h^{(2)} &\coloneqq -i \boldsymbol\sigma_2 = \begin{pmatrix} 0 & -1 \\ 1 & 0 \\ \end{pmatrix},
\end{split}
\end{align}
where $\boldsymbol\sigma_j$ is the $j$-th Pauli matrix. $h^{(1)}$ and $h^{(2)}$ generate the quaternion group $Q_8 = \Set{\pm \I, \pm i \boldsymbol\sigma_j | j=1,2,3}$ which has order eight. We obtain a representation $\varphi$ of $Q_8$ on $\Hi$ by setting
\begin{equation*}
\varphi(h) \coloneqq h_\otimes = h \otimes h \text{ for }h\in Q_8.
\end{equation*}
What are the irreducible subrepresentations that $\varphi$ decomposes into? The action of $\varphi(h^{(1)})$ and $\varphi(h^{(2)})$ on the Bell states is given by
\begin{align*}
\begin{split}
h^{(1)}_\otimes \ket{\Phi_\pm} &= \frac{1}{\sqrt{2}}(-\ket{1}\ket{1} \mp \ket{2}\ket{2}) = - \ket{\Phi_\pm} \\
h^{(1)}_\otimes \ket{\Psi_\pm} &= \frac{1}{\sqrt{2}}(\ket{1}\ket{2} \pm \ket{2}\ket{1}) = \ket{\Psi_\pm} \\
h^{(2)}_\otimes \ket{\Phi_\pm} &= \frac{1}{\sqrt{2}}(\ket{2}\ket{2} \pm \ket{1}\ket{1}) = \ket{\Phi_\pm} \\
h^{(2)}_\otimes \ket{\Psi_\pm} &= \frac{1}{\sqrt{2}}(-\ket{2}\ket{1} \mp \ket{1}\ket{2}) = -\ket{\Psi_\pm}.
\end{split}
\end{align*}
Thus the four one dimensional subspaces spanned by the Bell states are invariant under $\varphi(h^{(1)})$ and $\varphi(h^{(2)})$, i.e. $\varphi$ decomposes into four one dimensional subrepresentations. In particular these subrepresentations are pairwise inequivalent, as the action of $Q_8$ on distinct Bell states differs in signs. We now consider the random unitary operation
\begin{equation*}
\tilde{T}X \coloneqq \sum \limits_{i=1}^2 p_i h^{(i)}_\otimes X\ h^{(i)\dagger}_\otimes.
\end{equation*}
By Theorem \ref{thm:eigenspace_group} the eigenvectors of $\tilde{T}$ corresponding to eigenvalue 1 are the twirling states of $Q_8$, which are by Theorem \ref{thm:eigenvector_diagonal} diagonal in the Bell basis as $Q_8$ is finite (note that $\braket{A_1\otimes A_1,...,A_k \otimes A_k} \simeq \braket{A_1,...,A_k}$ for all unitary operators $A_1,...,A_k$). Thus we have achieved diagonality of the asymptotic eigenvectors of $\tilde{T}$, which is necessary for asymptotic preparation of Werner states, but not sufficient. It remains to identify the diagonal matrix elements corresponding to the Bell states $\ket{\Phi_\pm}$ and $\ket{\Psi_+}$. We therefore introduce another unitary operator $U$ which acts on $\Hi$ by permuting these states (up to some phase). 
\begin{equation*}
U \coloneqq \frac{1}{2} \begin{pmatrix}
i+1 & i+1 \\
i-1 & 1-i \\
\end{pmatrix}
\end{equation*}
Indeed, we find
\begin{align*}
\begin{split}
U_\otimes \ket{\phi_+} &= \frac{1}{\sqrt{2}} \left[ \left(\frac{i+1}{2}\ket{1} + \frac{i-1}{2}\ket{2}\right)^{\otimes 2} + \left(\frac{i+1}{2}\ket{1} + \frac{1-i}{2}\right)^{\otimes 2} \right] \\
&= \frac{i}{\sqrt{2}}( \ket{1}\ket{1} - \ket{2}\ket{2} ) \\
&= i \ket{\Phi_-}
\end{split}
\end{align*}
and analogously
\begin{align*}
\begin{split}
U_\otimes \ket{\Phi_-} &= - \ket{\Psi_+} \\
U_\otimes \ket{\Psi_+} &= i \ket{\Phi_+} \\
U_\otimes \ket{\Psi_-} &= \ket{\Psi_-}
\end{split}.
\end{align*}
We modify $\tilde{T}$ to obtain a random unitary operation
\begin{equation}
T X \coloneqq \tilde{T}X + p_3 U_\otimes X U_\otimes^\dagger,
\end{equation}
where, of course, the factors $p_1, p_2, p_3 \in (0,1)$ have to be adjusted to satisfy $p_1 +p_2+p_3 = 1$. Let $X \in \ker(T - \I)$. Then by the previous arguments above $X$ is Bell diagonal. Further Theorem \ref{thm:quantumop_eigenspace} implies that $X$ is an eigenvector of $U_\otimes . U_\otimes^\dagger$ as well, so in particular $X$ is an eigenvector of $U_\otimes^\dagger . U_\otimes$ for eigenvalue 1. It follows
\begin{equation*}
\bra{\Phi_+}X\ket{\Phi_+} = \bra{\Phi_+}U^\dagger_\otimes X U_\otimes \ket{\Phi_+} = \bra{\Phi_-}X\ket{\Phi_-} = \bra{\Phi_-}U^\dagger_\otimes X U_\otimes\ket{\Phi_-} = \bra{\Psi_+}X\ket{\Psi_+},
\end{equation*}
which shows that according to equation \eqref{eq:werner_2d} $X$ is a Werner state. As $X \in \ker(T - \I)$ was chosen arbitrarily we finally conclude that $\ker(T-\I) = \mathrm{span}(P_\mathrm{sym},P_\mathrm{asym})$, i.e. if $T$ is asymptotically stationary then $T^n X_0$ converges to a Werner state for all $X_0 \in \B$ and $n \to \infty$.
\subsection{Convergence Properties} \label{sec:numerical_prodedure}
\begin{figure}[h!]
\centering
\if\compileimages1
\begin{tikzpicture}
\begin{axis}[
    xlabel={$p$},
    ylabel={$\lambda_\mathrm{max}$},
    xmin=0, xmax=1,
    grid=both,
    height=0.5\textwidth,
    width=0.7\textwidth,
    /pgf/number format/.cd,
]
 
\addplot[
    color=black,
    ]
    table{maxeig_sl.txt};
\end{axis}
\end{tikzpicture}
\else
\includegraphics[width=0.8\textwidth]{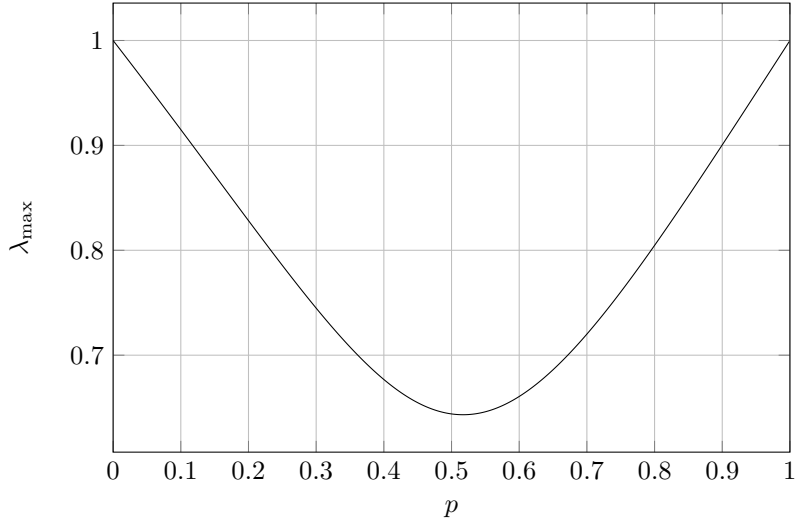}\hspace*{1.5cm}
\fi
\caption{Convergence rate of the random unitary operation associated with the generators $-i \boldsymbol\sigma_3$ and $i \boldsymbol\sigma_2 U^2$ of $\mathrm{SL}(2,3)$.}
\label{fig:maxeig_sl}
\end{figure}

We now define the group $G \coloneqq \braket{h_1, h_2, U} \subseteq \U{2}$, which turns out to be isomorphic to the group $\mathrm{SL}(2,3)$ of $2 \times 2$ matrices with determinant 1 over the field $\mathbb{Z}/3\mathbb{Z}$, which has order 24. It immediately follows from Theorem \ref{thm:eigenspace_group} that for any choice of operators $A_1,...,A_k \in G$ with $G = \braket{A_1,...,A_k}$ the corresponding random unitary operation \eqref{eq:random_unitary} with $U_i = A_i \otimes A_i$ has the same asymptotic property, i.e. if it is asymptotically stationary then it prepares Werner states asymptotically. A close analysis using the computer algebra system GAP showed that for the case of two generators, there are 192 pairs of operators that generate $G$, whereas only 144 of them are asymptotically stationary. Astonishingly, these 144 pairs are exactly the ones out of the 192 pairs for which the associated random unitary operation is diagonalizable, which is a fact that is yet to be investigated.
\\ \\
One can compare these pairs in terms of convergence rates of the resulting random unitary operation. Recall that for a diagonalizable random unitary operation the iterative behaviour is determined by its eigenvalues. According to the estimate \eqref{eq:estimate_diag} the quantity
\begin{equation*}
\lambda_\mathrm{max} = \mathrm{max} \{|\lambda|\ | \lambda \in \sigma \setminus \sigma_1\}
\end{equation*}
yields an appropriate value for comparing convergence rates. Of course, the value of $\lambda_\mathrm{max}$ depends on the choice of the coefficients $p_j$. In the case of two generators we have only one independent parameter $0<p<1$, as the other coefficient is given by $1-p$. Thus we can variate $p$ to minimize $\lambda_\mathrm{max}$.
\\ \\
A numerical evaluation reveals that in the case of two generators of $\mathrm{SL}(2,3)$ the optimal value is 
\begin{equation*}
\lambda_\mathrm{max} = 0.64324745(1),
\end{equation*}
which is obtained for instance\footnote{There are in total 96 pairs of generators yielding the same value.} by the generators $-i \boldsymbol\sigma_3$ and $i \boldsymbol\sigma_2 U^2$. Note that $U$ is an element of the normalizer\footnote{This can be checked easily using GAP.} of $Q_8$, i.e. for all $g \in Q_8$ there exists some $h \in Q_8$ such that $gU = Uh$. It follows that any element of SL(2,3) can be written as $gU^k$ with $g \in Q_8$ and $k=0,1,2$. Figure \ref{fig:maxeig_sl} shows $\lambda_\mathrm{max}$ as a function of $p$, we obtain a global minimum for $p = 0.51705601(1)$.

\begin{figure}[h!]
\centering
\if\compileimages1
\begin{tikzpicture} 
\begin{axis}[view={20}{45},
             xlabel=$p_1$,
             ylabel=$p_2$,
             colorbar,
             height=0.7\textwidth,
    	     width=0.8\textwidth]
    \addplot3[only marks,scatter,mark size=0.6] table {sl23_3d.txt};
  \end{axis}
\end{tikzpicture}
\else
\includegraphics[width=0.8\textwidth]{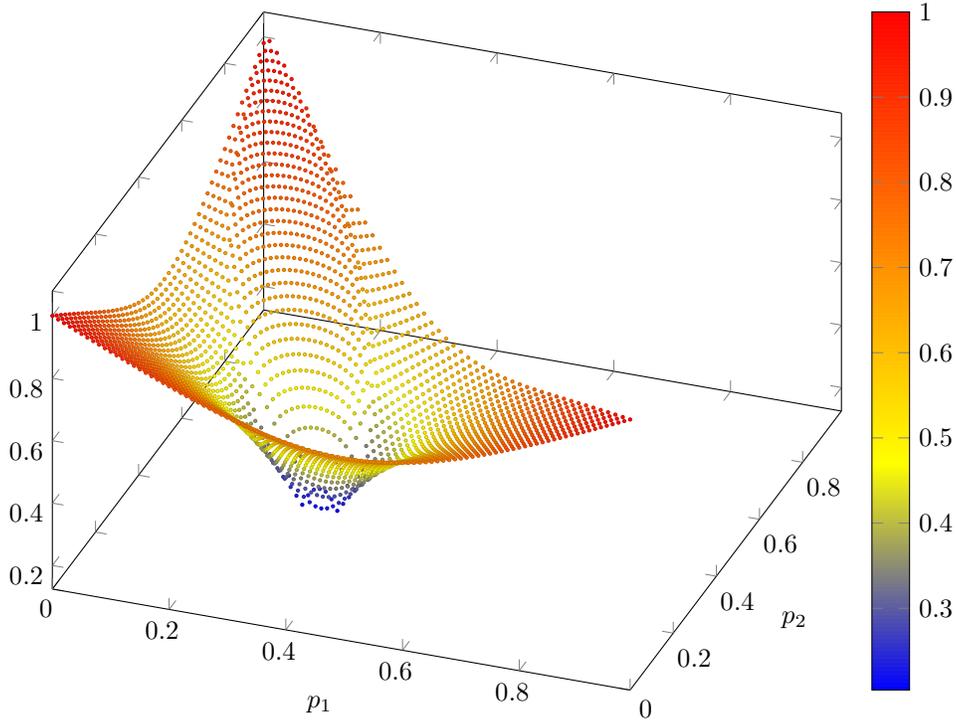}
\fi
\caption{Convergence rate of the random unitary operation associated to the generators $i\boldsymbol \sigma_3 U, -i \boldsymbol \sigma_3$ and $-i \boldsymbol \sigma_3 U^2$ of $\mathrm{SL}(2,3)$.}
\label{fig:sl23_3d}
\end{figure}

In the case of three generators there are 1888 triples that generate all of SL(2,3), whereas 112 of them are not asymptotically stationary. The numerical evaluation gets more involved in this case, since $\lambda_\mathrm{max}$ as a function of the parameters $p_1$ and $p_2$ is not differentiable everywhere, and the parameter variation underlies the constraint $p_1 + p_2 < 1$. Using a sequential least squares programming algorithm\cite{slsqp} it turns out that for different initial parameters the resulting minimized value $\lambda_\mathrm{max}$ varies in orders of magnitude $10^{-2}$. Since it is not the aim of this thesis to determine the very best value possible, but to provide a sufficiently accurate estimate for later comparisons, we consider a fixed number of minimized values of $\lambda_\mathrm{max}$ resulting from randomly chosen initial parameters. The approach that has been used for the results presented here goes as follows: 
\\ \\
At first a large number $N_\mathrm{ran}$ of pairs of parameters $(p_1, p_2)$ are generated in a uniform way that is compatible with the constraint $p_1 + p_2 < 1$. Then the corresponding value $\lambda_\mathrm{max}(p_1, p_2)$ is computed for each pair. The parameters resulting in the lowest value of $\lambda_\mathrm{max}$ are used as initial values for the sequential least squares programming algorithm, which yields a minimized value $\hat{\lambda}_\mathrm{max}(\hat{p}_1, \hat{p}_2)$. This procedure is repeated $N_\mathrm{opt}$ times, the best value of $\hat{\lambda}_\mathrm{max}(\hat{p}_1, \hat{p}_2)$ is the result.
\\ \\
Using this procedure, for $N_\mathrm{ran} = 300$ and $N_\mathrm{opt} = 50$ the best possible value that was found for three generators is $\hat{\lambda}_\mathrm{max} = 0.17157291(1)$, resulting from $i\boldsymbol \sigma_3 U, -i \boldsymbol \sigma_3$ and $-i \boldsymbol \sigma_3 U^2$ with 
\begin{equation*}
(\hat{p}_1, \hat{p}_2, \hat{p}_3) = (0.34314591, 0.31370849, 0.34314560).
\end{equation*}
Figure \ref{fig:sl23_3d} shows $\lambda_\mathrm{max}$ as a function of $p_1$ and $p_2$. The surface is piecewise smooth, but there are edges that probably originate from a change of the maximal eigenvalue. It looks like there is indeed a global minimum, it goes beyond the scope of this work though to prove that and determine the minimal value of $\lambda_\mathrm{max}$.

\section{Preparation of Werner States in Arbitrary Finite Dimensional Bipartite Systems}\label{ch:werner_dd}
We now consider the general case of a $d^2$ dimensional bipartite system for $d \geq 2$. Again we fix some orthonormal basis $(\ket{1},...,\ket{d})$ of $\Hi_\mathrm{A}$ (or $\Hi_\mathrm{B}$, respectively) and consider the orthonormal basis 
\begin{align}
\begin{split}
\ket{\phi_i} &\coloneqq \ket{i}\ket{i}, \\
\ket{\phi_{i,j}} &\coloneqq \frac{1}{\sqrt{2}}( \ket{i}\ket{j} + \ket{j}\ket{i} ) \\
\ket{\psi_{i,j}} &\coloneqq \frac{1}{\sqrt{2}}( \ket{i}\ket{j} - \ket{j}\ket{i} ),
\end{split}\tag{\ref{eq:werner_basis}}
\end{align}
of $\Hi$, w.r.t. which the projections \eqref{eq:projection_ops} are diagonal and take the form
\begin{align}
\begin{split}
P_\mathrm{sym} &= \sum \limits_i \ket{\phi_i}\bra{\phi_i} + \sum \limits_{i<j} \ket{\phi_{i,j}}\bra{\phi_{i,j}} \\
P_\mathrm{asym} &= \sum \limits_{i<j} \ket{\psi_{i,j}}\bra{\psi_{i,j}}. 
\end{split}\tag{\ref{eq:projections_d}}
\end{align}
In order to motivate a generalization, we recap the construction steps made in chapter \ref{ch:werner_2d}.
\begin{compactenum}[(1)]
\item{We found a finite group $H \subseteq \U{2}$ such that the representation $\varphi:H \to \GL{\Hi},U \mapsto U \otimes U$ decomposes into four inequivalent one dimensional subrepresentations, which are fortunately given by the Bell states, w.r.t. which the flip $F$ is diagonal. This guaranties that whenever some unitary Kraus operators generate $H$, the eigenvectors for eigenvalue 1 of the associated random unitary operation are Bell diagonal, which is neccessary for Werner states.}
\item{We found a unitary map $U \in \U{2}$ such that $U \otimes U$ permutes all basis vectors corresponding to $P_\mathrm{sym}$. This ensures that whenever some unitary Kraus operators generate $G = \braket{H, U}$, all eigenvectors for eigenvalue 1 of the associated quantum operation are Werner states. \\}
\end{compactenum}
One idea that comes up in order to generalize (1) is to consider cyclic groups, i.e. groups that are generated by a single element. As such they are in particular abelian, so that all irreducible subrepresentations are one dimensional by Schur's lemma. Unfortunately, since we are dealing with product operators of the form $h \otimes h$, there is no straightforward way of ensuring that all one dimensional subrepresentations are pairwise inequivalent (for instance, if $h$ is diagonal then there is no way of distinguishing $\ket{\phi_{i,j}}$ and $\ket{\psi_{i,j}}$ in terms of $h \otimes h$). However, we can still choose $h \in \U{d}$ in a way that as many subrepresentations are inequivalent as possible, and leave it to the other Kraus operators to compensate for this flaw.
\\ \\
As for generalizing (2), in the case $d > 2$ the situation becomes more complicated since $U \otimes U$ has to simultaneously fix the matrix elements corresponding to both all the $\ket{\phi_i}$, $\ket{\phi_{i,j}}$ and all the $\ket{\psi_{i,j}}$. In this general setting it is quite difficult to come up with an explicit form of such an operator. But as it turns out, it is possible to use a permutation that acts transitively on all single particle states\footnote{That is, for all $k,l \in \{1,...,d\}$ there is some $n \in \mathbb{Z}$ such that $\ket{k} = U^n \ket{l}$.} and mix up the remaining orbits using a third Kraus operator. 
\subsection{Construction of a Random Unitary Operation}\label{sec:huv}
In the following we investigate the random unitary operation $T$ defined by
\begin{equation}\label{eq:ruop_d}
T X \coloneqq p_1 h_\otimes X h_\otimes^\dagger + p_2 U_\otimes X U_\otimes^\dagger + p_3 V_\otimes X V_\otimes^\dagger
\end{equation}
where $h, U, V \in \U{d}$ are given as follows:
\begin{itemize}
\item{For all $k \in \{1,...,d\}$ we set
\begin{equation*}
h \ket{k} \coloneqq \exp \left( 2 \pi i \frac{2^{k-1}}{2^d}\right) \ket{k},
\end{equation*}
i.e. $h$ introduces a phase to every single particle state.
The matrix representation of $h$ w.r.t. $(\ket{1},...,\ket{d})$ is given by a diagonal matrix with entries $ \mathrm{e}^{i \pi 2^{1-d}},\mathrm{e}^{i \pi 2^{2-d}},...,-1$. As $h$ is diagonal with all diagonal entries being roots of unity, it is clearly unitary and has finite order. In particular, $h_\otimes$ is diagonal w.r.t. the basis \eqref{eq:werner_basis}, therefore the representation $\varphi_h : \braket{h} \to \GL{\Hi}, h \mapsto h_\otimes$ decomposes into one dimensional subrepresentations that are given by these basis vectors. We claim that only $d(d-1)/2$ of the possible $d(d-1)$ pairs of distinct subrepresentations are equivalent, namely the ones corresponding to the basis vectors $\ket{\phi_{i,j}}$ and $\ket{\psi_{i,j}}$ for $i < j$. This follows from the fact that any real number has a unique binary expansion, so that if $q = 2^{-i} + 2^{-j}$ then there is only one possible choice for $i$ and $j$ with $i<j$. We conclude, using the Theorems \ref{thm:eigenvector_diagonal} and \ref{thm:eigenspace_group}, that the only possible non-zero matrix elements of an eigenvector $X \in \ker(h_\otimes . h^\dagger_\otimes - \I)$ w.r.t. the basis \eqref{eq:werner_basis} are the diagonal entries as well as $\bra{\phi_{i,j}} X \ket{\psi_{i,j}}$ and $\bra{\psi_{i,j}} X \ket{\phi_{i,j}}$ for $i<j$. Note that here we used the fact that $h$ generates a finite group.
}
\item{
As mentioned above, we set
\begin{equation*}
U \ket{k} \coloneqq \ket{(k\ \mathrm{mod}\,2)+1},
\end{equation*}
for $k \in \{1,...,d\}$. Physically speaking, $U$ permutes all single particle states. By the same reasoning as in chapter \ref{ch:werner_2d}, $U_\otimes$ fixes different matrix elements according to its orbits on the basis \eqref{eq:werner_basis}: If we have an eigenvector $X \in \ker(U_\otimes . U_\otimes^\dagger - \I)$ and some basis vectors $\ket{\alpha_{1/2}}, \ket{\beta_{1/2}}$ with $U_\otimes\ket{\alpha_1} = \ket{\beta_1}$ and $U_\otimes\ket{\alpha_2} = \ket{\beta_2}$, then we find that
\begin{equation*}
\bra{\alpha_1} X \ket{\alpha_2} = \bra{\alpha_1}U_\otimes^\dagger X U_\otimes \ket{\alpha_2} = \bra{\beta_1} X \ket{\beta_2}.
\end{equation*}
The orbits of $U_\otimes$ acting on the basis \eqref{eq:werner_basis} are given by
\begin{align}\label{eq:orbits}
\begin{split}
\ket{\phi_i} &\mapsto \ket{\phi_{i+1}} \mapsto ... \mapsto \ket{\phi_1} \mapsto ... \mapsto \ket{\phi_i} \\
\ket{\phi_{i,j}} &\mapsto \ket{\phi_{i+1,j+1}} \mapsto ... \mapsto \ket{\phi_{i+d-j,d}} \mapsto \ket{\phi_{1,i+d-j+1}} \mapsto ... \mapsto \ket{\phi_{i,j}} \\
\ket{\psi_{i,j}} &\mapsto \ket{\psi_{i+1,j+1}} \mapsto ... \mapsto \ket{\psi_{i+d-j,d}} \mapsto \ket{\psi_{1,i+d-j+1}} \mapsto ... \mapsto \ket{\psi_{i,j}}.
\end{split}
\end{align}
Note that any orbit of $\ket{\phi_{i,j}}$ (and $\ket{\psi_{i,j}}$, respectively) has at least one representative of the form $\ket{\phi_{1,k}}$ (and $\ket{\psi_{1,k}}$ resp.) for some $k \in \{1,...,d\}$. 
}
\item{We consider a general unitary operator $A$ on $\mathrm{span}(\ket{1},\ket{2})$, given by its matrix representation
\begin{equation*}
A \coloneqq \mathrm{e}^{i \varphi} \begin{pmatrix}
\alpha & \beta \\
-\overline{\beta} & \overline{\alpha} \\
\end{pmatrix},
\end{equation*}
where $0 \leq \varphi < 2 \pi$, $\alpha, \beta \in \mathbb{C}$ with $|\alpha|^2 + |\beta|^2 = 1$. Note that any unitary $2 \times 2$ matrix can be expressed in this form, which can be shown using the fact that column and row vectors of a unitary matrix form an orthonormal basis. Now we set $V \coloneqq A \oplus \I_{d-2}$. Investigating the action of $V_\otimes$ on the representatives $\ket{\phi_1}$, $\ket{\phi_{1,k}}$ and $\ket{\psi_{1,k}}$ of the orbits of $U_\otimes$, we find
\begin{align*}
V_\otimes \ket{\phi_1} &= \mathrm{e}^{2i\varphi}(\alpha^2 \ket{\phi_1} + \overline{\beta}^2 \ket{\phi_2} - \sqrt{2} \alpha \overline{\beta} \ket{\phi_{1,2}}) \\
V_\otimes \ket{\phi_{1,2}} &= \mathrm{e}^{2i\varphi}(\sqrt{2} \alpha \beta \ket{\phi_1} - \sqrt{2} \overline{\alpha\beta} \ket{\phi_2} + (|\alpha|^2-|\beta|^2) \ket{\phi_{1,2}}) \\
V_\otimes \ket{\psi_{1,2}} &= \mathrm{e}^{2i \varphi} \ket{\psi_{1,2}} \\
V_\otimes \ket{\phi_{1,k}} &= \mathrm{e}^{i \varphi} (\alpha \ket{\phi_{1,k}} - \overline{\beta} \ket{\phi_{2,k}}) \\
V_\otimes \ket{\psi_{1,k}} &= \mathrm{e}^{i \varphi} (\alpha \ket{\psi_{1,k}} - \overline{\beta} \ket{\psi_{2,k}}) \\
\end{align*}
where $k > 2$.
}
\end{itemize}
\subsection{Asymptotic Preparation of Werner States}\label{sec:construction_asymptotics}
Using the definitions of $h,U$ and $V$, as well as their properties that we have collected so far, we can now prove that, in the case of $T$ being asymptotically stationary, $T$ prepares Werner states asymptotically. The asymptotic stationarity of $T$ is addressed inthe next section.
\begin{Thm}\label{thm:general_3g}
Let $T$ be the random unitary operation defined in equation \eqref{eq:ruop_d}, with $h$, $U$ and $V$ from above. If $\alpha \neq 0 \neq \beta$, then we have $\ker(T - \I) = \mathrm{span}(P_\mathrm{sym},P_\mathrm{asym} )$.
\end{Thm}
\begin{proof}
Let $X \in \ker(T - \I)$. Corollary \ref{cor:eigenspace_intersection} yields that $X$ is a simultaneous eigenvector of $h_\otimes . h_\otimes^\dagger$, $U_\otimes . U_\otimes^\dagger$ and $V_\otimes. V_\otimes^\dagger$. Therefore its only non-vanishing non-diagonal matrix elements w.r.t. the basis \eqref{eq:werner_basis} are $\bra{\phi_{i,j}} X \ket{\psi_{i,j}}$ and $\bra{\psi_{i,j}} X \ket{\phi_{i,j}}$ for $i<j$, and matrix elements that belong to the same orbit of $U_\otimes$ (as in equation \eqref{eq:orbits}) coincide. By equation \eqref{eq:projections_d} our goal is to show that $X$ is diagonal and that all diagonal matrix elements corresponding to $\ket{\phi_i}$, $\ket{\phi_{i,j}}$, and respectively $\ket{\psi_{i,j}}$, coincide.
\\ \\
First we show that all non-diagonal matrix elements vanish. It holds
\begin{align*}
\bra{\phi_{1,2}}X\ket{\psi_{1,2}} = \bra{\phi_{1,2}} V_\otimes^\dagger X V_\otimes \ket{\psi_{1,2}} = \underbrace{(|\alpha|^2-|\beta|^2)}_{\neq 1}\bra{\phi_{1,2}}X\ket{\psi_{1,2}} = 0
\end{align*}
where we used $\alpha \neq 0 \neq \beta$. Moreover, we have for all $k > 1$
\begin{align*}
\bra{\phi_{1,k}}X\ket{\psi_{1,k}} &= \bra{\phi_{1,k}}V_\otimes^\dagger X V_\otimes \ket{\psi_{1,k}} = |\alpha|^2 \bra{\phi_{1,k}}X\ket{\psi_{1,k}} + |\beta|^2 \bra{\phi_{2,k}}X\ket{\psi_{2,k}} \\
&= \underbrace{\frac{|\beta|^2}{1-|\alpha|^2}}_{=1} \bra{\phi_{2,k}}X\ket{\psi_{2,k}} \\
&= \bra{\phi_{1,k-1}}X\ket{\psi_{1,k-1}}.
\end{align*}
By induction over $k$ it follows $\bra{\phi_{1,k}}X\ket{\psi_{1,k}} = 0$ for all $k>1$. Analogously one shows $\bra{\psi_{1,k}}X\ket{\phi_{1,k}} = 0$ for all $k>1$. Now we consider the diagonal matrix elements. We find
\begin{align*}
\bra{\phi_{1,2}}X\ket{\phi_{1,2}} &= \bra{\phi_{1,2}}V_\otimes^\dagger X V_\otimes\ket{\phi_{1,2}}= 2|\alpha \beta|^2( \braket{\phi_1|X|\phi_1} + \underbrace{\braket{\phi_2|X|\phi_2}}_{=\braket{\phi_1|X|\phi_1}} ) + (|\alpha|^2-|\beta|^2)^2 \braket{\phi_{1,2}|X|\phi_{1,2}} \\
&= \underbrace{\frac{4|\alpha \beta|^2}{1-(|\alpha|^2-|\beta|^2)^2}}_{=1} \bra{\phi_1}X\ket{\phi_1}
\end{align*}
Further we find for all $k > 2$
\begin{align*}
\bra{\phi_{1,k}}X\ket{\phi_{1,k}} &= \bra{\phi_{1,k}}V_\otimes^\dagger X V_\otimes \ket{\phi_{1,k}} = |\alpha|^2 \bra{\phi_{1,k}}X\ket{\phi_{1,k}} + |\beta|^2 \bra{\phi_{2,k}}X\ket{\phi_{2,k}} \\ 
&= \underbrace{\frac{|\beta|^2}{1-|\alpha|^2}}_{ = 1} \bra{\phi_{2,k}}X\ket{\phi_{2,k}} = \bra{\phi_{1,k-1}}X\ket{\phi_{1,k-1}}.
\end{align*}
Analogously it follows $\bra{\psi_{1,k}}X\ket{\psi_{1,k}} = \bra{\psi_{1,k-1}}X\ket{\psi_{1,k-1}}$, so by induction over $k$ we deduce 
\begin{equation*}
\bra{\phi_{1,k}}X\ket{\phi_{1,k}} = \bra{\phi_1}X\ket{\phi_1} \text{ and } \bra{\psi_{1,k}}X\ket{\psi_{1,k}} = \bra{\psi_{1,2}}X\ket{\psi_{1,2}}
\end{equation*}
for all $k > 1$, which completes the proof.
\end{proof}
Some remarks on the construction should be emphasized at this point.
\begin{itemize}
\item{Strictly speaking, the construction of $T$ is not a generalization of the random unitary operations considered in chapter \ref{ch:werner_2d}. This is actually an advantage since we can now compare the resulting random unitary operations in terms of convergence properties.}
\item{Obviously, the choices of $h, U$ and $V$ are not the most general ones allowing for the arguments used in the proof of Theorem \ref{thm:general_3g}. For the sake of simplicity, many arbitrary choices have been made during the construction. For instance, other choices for the diagonal entries of $h$ are possible, $U$ could be replaced by any unitary operator that acts transitively on all single particle states up to a phase, and in the construction of $V$ the matrix $A$ could have been defined on any other two dimensional subspace of $\Hi_\mathrm{A/B}$. However, for the purpose of this thesis we will stick to the choices of $h, U$ and $V$ from above.}
\item{In general, the group generated by $h,U$ and $V$ is of infinite order, so there is no easy way of determining all tuples of elements that generate the whole group, as in the case of SL(2,3). Nevertheless, there are lots of triples of elements for which it is obvious that they generate the same group, for instance we have 
\begin{equation*}
\braket{V^{\kappa_1}hV^{\kappa_2}, V^{\kappa_3}UV^{\kappa_4}, V} = \braket{h, U, V}
\end{equation*}
for all choices of $\kappa_1,...,\kappa_4 \in \mathbb{Z}$, and of course the roles of $h, U$ and $V$ are interchangeable. By Theorem \ref{thm:eigenspace_group} all of these triples have an associated random unitary operation that prepares Werner states asymptotically, if it is asymptotically stationary. Note that at this point we need the fact that Theorem \ref{thm:eigenspace_group} holds for infinite groups, which was the purpose of introducing the Haar measure in the first place.
}
\end{itemize}
\subsection{Asymptotic Stationarity} \label{sec:asymptotic_stationarity}
We found a random unitary operation $T$ depending on six parameters (four independent parameters for $\varphi$, $\alpha$ and $\beta$ and two independent parameters $p_1, p_2$) that prepares Werner states asymptotically, if it is asymptotically stationary. How do we have to choose these parameters in order to make $T$ asymptotically stationary? By Corollary \ref{cor:eigenspace_intersection} the set of eigenvalues with unit modulus $\sigma_1$ is independent of the choice of the $p_j$. Using the fact that we can investigate the asymptotic properties of a random unitary operation by looking at the individual Kraus operators once again, we obtain a useful result about the eigenvalues with unit modulus.
\begin{Cor} \label{cor:eigenvalue_product}
Let $T$ be a random unitary operation as in equation \eqref{eq:random_unitary} and $\lambda \in \sigma_1$. Then for all $m \in \{1,...,m\}$ there are some eigenvalues $\mu_1$ and $\mu_2$ of $U_m$ such that $\lambda = \mu_1 \overline{\mu_2}$.
\end{Cor}
\begin{proof}
Let $0 \neq X \in \ker( T - \lambda \I)$. Then by Corollary \ref{cor:eigenspace_intersection} we have $U_m X U_m^\dagger = \lambda X$ and therefore $U_m^\dagger X U_m = \overline{\lambda} X$. Since $U_m$ is unitary, it is diagonalizable (note that unitary operators are normal), so let $(\ket{b_1},...,\ket{b_n})$ be an eigenbasis of $U_m$ and $\mu_1,...,\mu_n$ be the associated eigenvalues. For all $i,j \in \{1,...,n\}$ we find
\begin{equation*}
\overline{\lambda} \bra{b_i}X\ket{b_j} = \bra{b_i} U_m^\dagger X U_m \ket{b_j} = \overline{\mu_i} \mu_j \bra{b_i}X\ket{b_j}.
\end{equation*}
Since $X \neq 0$, there is at least one non-vanishing matrix element, which yields the claim.
\end{proof}
Applying Corollary \ref{cor:eigenvalue_product} and the fact that both $h$ and $V$ are diagonalizable, we find that any eigenvalue $\lambda \in \sigma_1$ of $T$ can be written as $\lambda = \lambda_1 \lambda_2 \overline{\lambda_3 \lambda_4} = \mu_1 \mu_2 \overline{\mu_3 \mu_4}$ for eigenvalues $\lambda_i$ of $h$ and $\mu_i$ of $V$. The eigenvalues of $h$ and $V$ are given by
\begin{align*}
\sigma(h) &= \Set{ \mathrm{e}^{i \pi 2^{k-d}} | k=1,...,d} \\
\sigma(V) &= \Set{1, \mathrm{e}^{i(\varphi \pm \theta)}},
\end{align*}
where $\theta \in [0, 2\pi)$ is defined by $\cos(\theta) = \mathfrak{R}(\alpha)$. We conclude that $\sigma_1 \subseteq \{1\}$, and therefore asymptotic stationarity of $T$, is determined only by the choices of $\varphi$ and $\mathfrak{R}(\alpha)$. Moreover, using a simple argument about the eigenvalues of $h$ and $U$, we get the following result about $T$:
\begin{Obs}
The random unitary operation $T$ from section \ref{sec:huv} is asymptotically stationary for almost all choices of $A \in \mathcal{U}(d)$, i.e. the set of real parameters that determine $A$, for which $T$ is not guaranteed to be asymptotically stationary, has measure zero.
\end{Obs}
\begin{proof}
As we have that $\sigma_1 \subseteq \{ \lambda_1 \lambda_2 \overline{\lambda_3} \overline{\lambda_4}\ |\ \lambda_i \in \sigma(h) \}$, we conclude that all asymptotic eigenvalues of $T$ have to be rational roots of unity. Using the eigenvalues of $V$, we find
\begin{equation*}
\sigma_1 \subseteq \{ \lambda_1 \lambda_2 \overline{\lambda_3} \overline{\lambda_4}\ |\ \lambda_i \in \sigma(V) \} \subseteq \left\{ \exp\left( i(k \varphi + l \theta)\right)\ |\ k,l \in \{-4,...,4 \} \right\}.
\end{equation*}
Thus, if $\frac{k \varphi + l \theta}{\pi}$ is not one of the finitely many rational numbers that correspond to the possible rational root eigenvalues of $T$ for all $k, l \in \{ -4,...,4\}$, then 1 is the only asymptotic eigenvalue, i.e. $T$ is asymptotically stationary. Since these rational numbers form a set of measure zero within the reals and we are only concerned about finitely many values, this implies the statement.
\end{proof}
In particular, asymptotic stationarity will not be an issue in numerical computations, as the parameters can always be modified slightly to ensure convergence. Nevertheless, in the proof we only used the fact that the eigenvalues of $h_\otimes\, .\,h_\otimes^\dagger$ are rational roots of unity, which ignores some information we actually have about the eigenvalues. We also do not have to consider $\frac{k \varphi + l \theta}{\pi}$ for all $k, l \in \{ -4,...,4\}$, a combinatorical analysis reveales that actually considering $k$ and $l$ with $|k|+|l| \in \{2,4\}$ and $|k|\leq2$ suffices.
\\ \\
For any particular dimension $d$ a stronger sufficient condition for asymptotic stationarity of the random unitary operation $T$ depending on the parameters $\varphi$ and $\theta$ can be derived, using the arguments outlined above. As an example, let us consider the case of qubits here ($d=2$). In this case, the eigenvalues of the unitary operator $U$ from above are 1 and -1, which already yields $\sigma_1 \subseteq \{ \pm 1 \}$ and thus $-1 = \mathrm{e}^{i\pi}$ is the only possible eigenvalue that could potentially contribute to nonstationary asymptotic dynamics. Now, in qubit systems we have that $V = A$ and it is straightforward to show that the eigenvalues of $V_\otimes\,.\,V_\otimes^\dagger$ are given by
\begin{equation*}
\sigma(V_\otimes\,.\,V_\otimes^\dagger) = \Set{1, \mathrm{e}^{2i\varphi}, \mathrm{e}^{2i\theta}, \mathrm{e}^{\pm 4i\varphi}}.
\end{equation*}
Therefore, a sufficient condition for $T$ being asymptotically stationary in qubit systems is given by $\frac{\pi}{2} \notin \{ \varphi, \theta, \pm 2 \varphi\}$.
\subsection{Reduction to Two Generators}
What is the minimal number of Kraus operators of a random unitary operation required to prepare Werner states asymptotically? Using only one Kraus operator does not work, as in this case the resulting quantum operation is itself unitary. Thus, starting with a pure initial state will result after each iteration step in a pure state as well, which cannot converge to a Werner state as these states are not pure. But it cannot work for all other states either, since, in the case of an asymptotically stationary quantum operation, $T$ being unitary implies $\sigma = \sigma_1 \subseteq \{1\}$, showing that $T$ is in fact trivial. 
\\ \\
Consider the case $d = 2$ for the particular choice 
\begin{equation*}
A = \frac{1}{\sqrt{2}}\begin{pmatrix}
1 & 1 \\
1 & -1 \\
\end{pmatrix},
\end{equation*}
then we find using GAP that $\braket{h, U, V} = \braket{h,UV}$, which is a finite group of order 192 and has SL(2,3) as a normal subgroup. Thus the random unitary operation associated with the two generators $h$ and $UV$ prepares Werner states asymptotically as well. Astonishingly, this seems to be true in any dimension and without the restriction on $A$! In the case of the dimension $d$ of both subsystems being an odd number we can prove this and obtain the following result:
\begin{Thm}\label{thm:general_2g}
Let the dimension $d$ be odd, let $\tilde{T}$ be the random unitary operation associated with the unitary Kraus operators $h_\otimes$ and $(UV)_\otimes$, with $h, U$ and $V$ from section \ref{sec:huv} and $\alpha \neq 0 \neq \beta$. Then we have $\ker(\tilde{T} - \I) = \mathrm{span}(P_\mathrm{sym},P_\mathrm{asym} )$.
\end{Thm}
The proof is presented in appendix \ref{sec:general_2g_proof}. While the special case $d = 2$ for the specific choice for $A$ as above follows from a beautiful group theory argument, and even the arbitrary dimensional case for three generators was fairly easy, the proof of Theorem \ref{thm:general_2g} is considerably harder. This is due to the fact that even though most of the crucial arguments used in the three generator case stay valid, the calculations involve a complicated arbitrary dimensional system of linear equations. Numerical evidence suggests that the Theorem holds true in any dimension, though all attempts to prove this for the purpose of this thesis were not successful. Note that the construction of two generators works in dimension 2 as well, which is very easy to check and therefore not carried out here.
\subsection{Further Ideas}\label{sec:further_ideas}
At this point one might ask wether having a minimal number of Kraus operators to prepare is even of practical relevance, since it might as well be the case that preparing more Kraus operators is easier practically and yields better convergence rates. A detailed investigation of this problem goes beyond the scope of this thesis, here we simply assume that any unitary operator can be implemented practically, which at least in qubit systems is a reasonable assumption \cite{quantuminformationscience}. However, here are some further ideas supposing that the restriction of having a minimal number of Kraus operators is being dropped.
\begin{itemize}
\item{Adding the identity operator $\I$ as unitary Kraus operator ensures asymptotic stationarity of the resulting random unitary operation. This is due to Corollary \ref{cor:eigenvalue_product}, using the fact that the only eigenvalue of the identity is 1.}
\item{Given a random unitary operation $T$ with unitary Kraus operators $U_1,...,U_k$ and factors $p_1,...,p_k$, one can construct a random unitary operation that is guarantied to be diagonalizable unitarily. Simply set
\begin{equation}
\tilde{T}(X) \coloneqq \frac{1}{2} \sum \limits_{i = 1}^k p_i (U_i X U_i^\dagger + U_i^\dagger X U_i)
\end{equation}
for $X \in \B$ and observe that $\tilde{T}$ is self-adjoint, whence it can be diagonalized in terms of a unitary operator. This makes the estimate \ref{eq:estimate_diag_onb} valid, thus this construction might turn out to be usefull whenever an explicit convergence rate bound is needed. Maybe there is even a clever practical way of preparing the operator $U_i^\dagger$ if the operator $U_i$ is prepared already.
}
\end{itemize}
\section{Application of the Construction in Qubit and Qudrit Systems}\label{ch:application}
In the previous chapter we established a general construction of random unitary operations that prepare Werner states asymptotically. Now we aim to apply this result in qubit ($d=2$) and qudrit ($d=3$) systems and discuss convergence issues. In general we have six independent parameters (four for $A$ and two for the coefficients $p_1$, $p_2$) to variate in order to minimize $\lambda_\mathrm{max}$. So for finding suitable random unitary operations with optimal convergence properties one has a variation problem at hand, which can be handled numerically.
\\ \\
As for the optimization of $\lambda_\mathrm{max}$, we make use of the fact described in section \ref{sec:construction_asymptotics} that 
\begin{equation*}
\braket{V^{\kappa_1}hV^{\kappa_2}, V^{\kappa_3}UV^{\kappa_4}, V} = \braket{h, U, V}
\end{equation*}
for all $\kappa_1,...,\kappa_4 \in \mathbb{Z}$. In order to determine the best choice systematically, we can interpret $\kappa_1,...,\kappa_4$ as the components of a multiindex $\kappa \in \mathbb{Z}^4$ and categorize it by its 1-norm 
\begin{equation}
|\kappa| \coloneqq \sum \limits_{i=1}^4 |\kappa_i| \in \mathbb{N}_0.
\end{equation}
Despite defining a partial order on the set of multiindices, for this particular problem $|.|$ has the advantage of measuring the amount of matrix multiplications in order to obtain $V^{\kappa_1}hV^{\kappa_2}$ and $V^{\kappa_1}UV^{\kappa_2}$ from $h$ and $U$. Of course, the roles of $h, U$ and $V$ here are interchangeable, and in the case of two generators one can consider multiindices in $\mathbb{Z}^2$ analogously. For the purpose of this thesis all random unitary operations corresponding to multiindices $\kappa$ up to $|\kappa| \leq 3$ have been investigated using the procedure described in section \ref{sec:numerical_prodedure}, with $N_\mathrm{ran} = 300$ and $N_\mathrm{opt} = 50$. In order to ensure asymptotic stationarity during the numerical analysis, $\lambda_\mathrm{max}$ is set to 1 if there are more that two eigenvalues with unit modulus (counted with multiplicity). As stated in section \ref{sec:asymptotic_stationarity}, this can only be the case for a few discrete choices of parameters (if at all), so the restriction should have no significant impact on the numerical variation.
\subsection{Results of a Numerical Optimization} \label{sec:application}
\begin{figure}[h!]
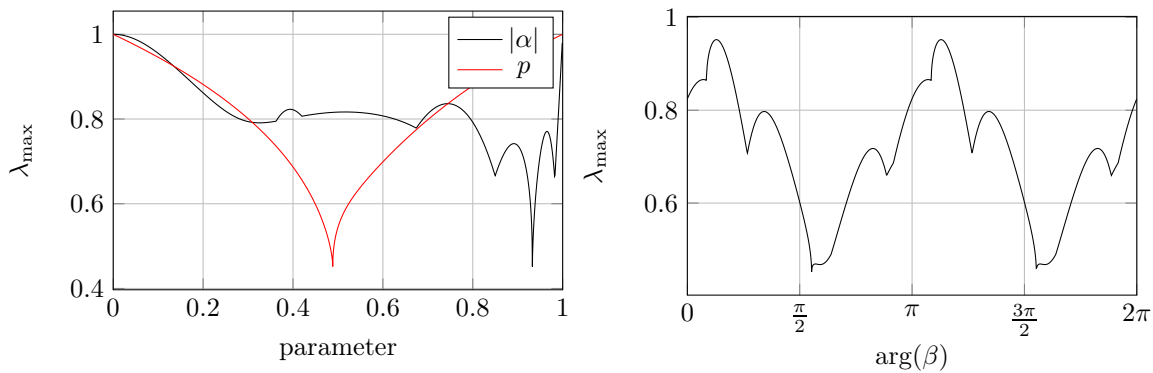
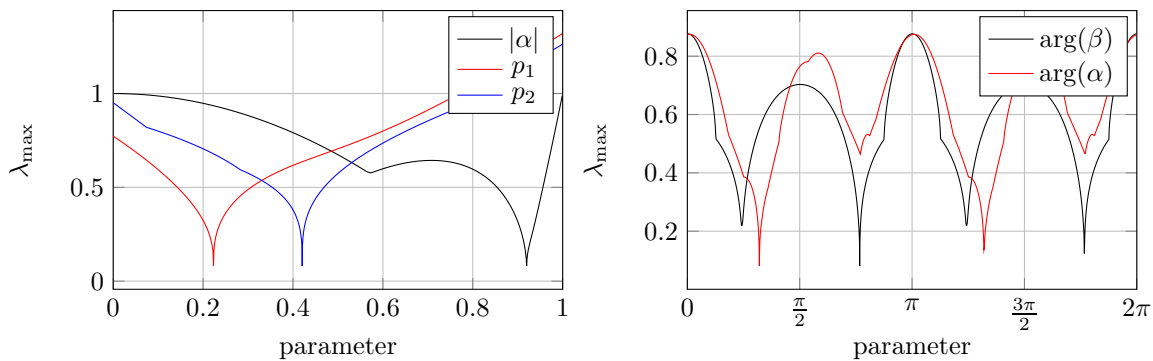

\centering
\begin{subfigure}{1.0\textwidth}
\if\compileimages1
\begin{minipage}{0.5\textwidth}
\begin{tikzpicture}
\begin{axis}[
    xlabel={parameter},
    ylabel={$\lambda_\mathrm{max}$},
    xmin=0, xmax=1,
    grid=both,
    width=\textwidth,
    height=0.7\textwidth,
    /pgf/number format/.cd,
]
 
\addplot[color=black]
    table[x=x,y=|alpha|]{maxeig_d2g2_prange.txt};
\addlegendentry{$|\alpha|$}
\addplot[color=red]
    table[x=x,y=p1]{maxeig_d2g2_prange.txt};
\addlegendentry{$p$}
\end{axis}

\end{tikzpicture}
\end{minipage}
\begin{minipage}{0.5\textwidth}
\begin{tikzpicture}
\begin{axis}[
    xlabel={$\arg(\beta)$},
    ylabel={$\lambda_\mathrm{max}$},
    xmin=0, xmax=2*pi,
    xtick={0, 1.570796, 3.141593, 4.712389, 6.28319},
    xticklabels={0, $\frac{\pi}{2}$, $\pi$, $\frac{3\pi}{2}$, $2\pi$},
    grid=both,
    width=\textwidth,
    height=0.7\textwidth,
    /pgf/number format/.cd,
]
 
\addplot[color=black]
    table[x=x,y=arg(beta)]{maxeig_d2g2_phirange.txt};
\end{axis}
\end{tikzpicture}
\end{minipage}
\else
\includegraphics[width=0.5\textwidth]{maxeig_d2g2_prange}
\includegraphics[width=0.5\textwidth]{maxeig_d2g2_phirange}
\fi
\subcaption{The convergence parameter $\lambda_\mathrm{max}$ of the random unitary operation associated with the generators $(V^\dagger U^\dagger)^3h$ and $UV$. A variation of the missing parameters $\varphi$ and $\arg(\alpha)$ leaves $\lambda_\mathrm{max}$ unchanged, which are therefore omitted.}
\label{fig:maxeig_d2g2}
\vspace{1cm}
\end{subfigure}
\begin{subfigure}{1.0\textwidth}
\if\compileimages1
\begin{minipage}{0.5\textwidth}
\begin{tikzpicture}
\begin{axis}[
    xlabel={parameter},
    ylabel={$\lambda_\mathrm{max}$},
    xmin=0, xmax=1,
    grid=both,
    width=\textwidth,
    height=0.7\textwidth,
    /pgf/number format/.cd,
]
 
\addplot[color=black]
    table[x=x,y=|alpha|]{maxeig_d2g3_prange.txt};
\addlegendentry{$|\alpha|$}
\addplot[color=red]
    table[x=x,y=p1]{maxeig_d2g3_prange.txt};
\addlegendentry{$p_1$}
\addplot[color=blue]
    table[x=x,y=p2]{maxeig_d2g3_prange.txt};
\addlegendentry{$p_2$}
\end{axis}
\end{tikzpicture}
\end{minipage}
\begin{minipage}{0.5\textwidth}
\begin{tikzpicture}
\begin{axis}[
    xlabel={parameter},
    ylabel={$\lambda_\mathrm{max}$},
    xmin=0, xmax=2*pi,
    xtick={0, 1.570796, 3.141593, 4.712389, 6.28319},
    xticklabels={0, $\frac{\pi}{2}$, $\pi$, $\frac{3\pi}{2}$, $2\pi$},
    grid=both,
    width=\textwidth,
    height=0.7\textwidth,
    /pgf/number format/.cd,
]
 
\addplot[color=black]
    table[x=x,y=arg(beta)]{maxeig_d2g3_phirange.txt};
\addlegendentry{$\arg(\beta)$}
\addplot[color=red]
    table[x=x,y=arg(alpha)]{maxeig_d2g3_phirange.txt};
\addlegendentry{$\arg(\alpha)$}
\end{axis}
\end{tikzpicture}
\end{minipage}
\else
\includegraphics[width=0.5\textwidth]{maxeig_d2g3_prange}
\includegraphics[width=0.5\textwidth]{maxeig_d2g3_phirange}
\fi
\subcaption{The convergence parameter $\lambda_\mathrm{max}$ of the random unitary operation associated with the generators $hV,U(V^\dagger)^2$ and $V$. Once again, variation of $\varphi$ has no effect on $\lambda_\mathrm{max}$.}
\label{fig:maxeig_d2g3}
\end{subfigure}
\caption{The convergence parameter $\lambda_\mathrm{max}$ of the random unitary operation associated with different tuples of generators in a neighbourhood of the optimization minimum from table \ref{tab:params_2d} as a one dimensional function of each individual parameter.}
\end{figure}

First we consider the case of two generators. The best value that was found is $\lambda_\mathrm{max,2g} = 0.45171364(1)$, resulting from the generators $(V^\dagger U^\dagger)^3h$ and $UV$, where all parameters are specified in table \ref{tab:params_2d}. This value is significantly better than the one obtained from generators of SL(2,3), which was $\lambda_\mathrm{max,2g,SL} = 0.64324745(1)$. Since we are dealing with 5 parameters we cannot visualize $\lambda_\mathrm{max}$ as a function of all its parameters. We can however fix all parameters but one, to get an idea of what the neighbourhood of the optimization minimum looks like. The result is depicted in figure \ref{fig:maxeig_d2g2}. As expected we get piecewise smooth functions with edges at the points where the eigenvalue with maximal absolute value switches. A variation of $\arg(\alpha)$ or $\varphi$ has no effect on $\lambda_\mathrm{max}$, so we are free to choose any value\footnote{The values of $\arg(\alpha)$ and $\varphi$ in table \ref{tab:params_2d} are the output of the sequential least squares algorithm and have no further meaning.}.
\\ \\
In the case of three generators the best value that was found is $\lambda_\mathrm{max,3g} = 0.08044643(1)$, resulting from the generators $hV, U(V^\dagger)^2$ and $V$ with the parameters in table \ref{tab:params_2d}. Once again the value is significantly better than the one obtained from SL(2,3), which was $\lambda_\mathrm{max,3g,SL} = 0.17157291(1)$. Analogously to the two generator case, figure \ref{fig:maxeig_d2g3} shows $\lambda_\mathrm{max}$ as a function of each individual parameter in the neighbourhood of the minimum. Once again $\lambda_\mathrm{max}$ is independent of $\varphi$, though now the value of $\arg(\alpha)$ has a significant impact. 
\\ \\
Now we consider bipartite qudrit systems. For two generators we have the best value $\lambda_\mathrm{max,2g} = 0.67402461(1)$ associated with the generators $h$ and $UV$, while for three generators we have $\lambda_\mathrm{max,3g} = 0.55847203(1)$ associated with the generators $h$, $UV$ and $U$. The parameters are given in table \ref{tab:params_2d}. Note that the values for $\lambda_\mathrm{max}$ are considerably higher in three than in two dimensions. The figures \ref{fig:maxeig_d3g2} and \ref{fig:maxeig_d3g3} in appendix \ref{ch:numerical_appendix} show $\lambda_\mathrm{max}$ as a one-dimensional function of each individual parameter, but there is actually nothing new there. The appendix also contains a table (table \ref{tab:params_top10}) with details about the best 10 random unitary operations of each category, which the numerical evaluations have revealed.

\begin{table}[!htb]
\begin{center}
 \caption{Parameters of the random unitary operations resulting in the best value for $\lambda_\mathrm{max}$ in dimensions two and three.}
 \label{tab:params_2d}
 \scalebox{0.9}{
 \begin{tabular}{c c c c c c c c c}
 \hline \hline
  $d$ & generators & $\lambda_\mathrm{max}$ & $\varphi$ & $|\alpha|$ & $\arg( \alpha )$ & $\arg( \beta )$ & $p_1$ & $p_2$ \\ \hline
  2 & $(V^\dagger U^\dagger)^3 h,UV$ & 0.45171364(1) & 0.02040734 & 0.93254821 & 2.89447138 & 4.87953210 & 0.48788387 & - \\
  2 & $hV,U (V^\dagger)^2,V$ & 0.08044643(1) & 4.18274534 & 0.91986392 & 1.00676113 & 2.40909110 & 0.22305894 & 0.42044878 \\
  3 & $h,UV$ & 0.67402461(1) & 2.86002806 & 0.74921865 & 3.66908666 & 2.32545709 & 0.49097422 & - \\
  3 & $h,UV,U$ & 0.55847203(1) & 0.36470858 & 0.58546063 & 4.94276220 & 1.53168497 & 0.35903090 & 0.41473933 \\
  \hline \hline
 \end{tabular}}
\end{center}
\end{table}
\newpage
\subsection{Numerical Example}
In the previous section we optimized the secound greatest modulus of the eigenvalues of a random unitay operation as a measure for the convergence rate of its iteration. However, according to section \ref{sec:quantum_op_asymp} this is technically only a valid method for diagonalizable random unitary operations. In principle diagonalizability of a linear map is easily checked by computing the geometric multiplicities of each individual eigenvalue\footnote{Recall that a linear map is diagonalizable if and only if all the geometric multiplicities add up to the dimension.}. Unfortunately, because of limitations due to finite machine precision it is in some cases not possible to tell whether a given linear map is diagonalizable in numerical applications. In the following we consider the random unitary operations from section \ref{sec:application} with optimized parameters and investigate their rate of convergence as a numerical example.
\\ \\
If the random unitary operation $T$ prepares Werner states, then by Proposition \ref{obs:werner_projection} we have that $T^n \to \mathcal{P}$ for $n \to \infty$, where
\begin{equation}\label{eq:twirling_op}
\mathcal{P} \coloneqq \mathfrak{t}_\mathcal{U} = \sqrt{\frac{2}{d(d+1)}} \braket{P_\mathrm{sym},.}_\mathrm{HS} + \sqrt{\frac{2}{d(d-1)}} \braket{P_\mathrm{asym},.}_\mathrm{HS}
\end{equation}
is the twirling operation that projects any state to its corresponding Werner state. Since in the case of random unitary operations dual eigenvectors coincide with eigenvectors, we can strengthen equation \eqref{eq:estimate_diag} to obtain an explicit estimate for the convergence of the iteration of $T$, assuming that $T$ is diagonalizable: By equation \eqref{eq:estimate_diag_onb} in the case of random untiary operations we have that $\|P_\lambda\| \leq d_\lambda$ for every eigenprojection $P_\lambda$, where $d_\lambda$ is the geometric multiplicity of the eigenvalue $\lambda \in \sigma$. Thus, for all $X \in \B$ and $n \in \mathbb{N}$ we find
\begin{align*}
\|T^n X - \mathcal{P}(X) \| &\leq \sum \limits_{\lambda \in \sigma \setminus \sigma_1} |\lambda|^n \| P_\lambda(X) \| \\
&\leq \lambda_\mathrm{max}^n \cdot \|X\| \sum \limits_{\lambda \in \sigma \setminus \sigma_1} d_\lambda \\
&\leq (d^4 - 2)\lambda_\mathrm{max}^n \cdot \|X\|.
\end{align*}
Here we used the fact that $\sigma_1 = \{1\}$ by asymptotic stationarity of $T$, and $d_1 = 2$ by Theorem \ref{thm:werner}.
This shows the estimate $\|T^n -\mathcal{P}\|_\mathcal{B} \leq  (d^4 - 2)\lambda_\mathrm{max}^n$ in the operator norm $\|.\|_\mathcal{B}$ w.r.t. the Hilbert Schmidt norm on $\B$. This norm is, however, not easy to compute practically, which is why we consider the Hilbert Schmidt norm $\|.\|_\mathrm{HS}$ instead and use the norm equivalence $\|A\|_\mathcal{B} \leq \|A\|_\mathrm{HS} \leq \sqrt{n} \|A\|_\mathcal{B}$ for all $A \in \mathbb{C}^{(n,n)}$. Together we obtain the estimate
\begin{equation}\label{eq:estimate_RUO}
\|T^n -\mathcal{P}\|_\mathrm{HS} \leq d^2(d^4-2) \lambda_\mathrm{max}^n,
\end{equation}
which is valid for all $n \in \mathbb{N}$ and under the assumption that $T$ is diagonalizable.
\\ \\
The subsequent figures \ref{fig:convergence_2d} and \ref{fig:convergence_3d} illustrate the convergence of the iterated random unitary operations from section \ref{sec:huv} with the parameters from table \ref{tab:params_2d} for qubits and qudrits. Plotted is the distance between the iterated quantum operation $T^n$ and the twirling operation $\mathcal{P}$ as a function of the number of iterations $n$, measured in the Hilbert Schmidt norm. Additionally, the continuous lines represent the bound \eqref{eq:estimate_RUO} for comparison. In the case of qubits (figure \ref{fig:convergence_2d}), the calculated points exceed the bound, which reveals that in fact the underlying random unitary operation is not diagonalizable. For qudrits on the other hand (figure \ref{fig:convergence_3d}), the bound is valid with quite a big gap, demonstrating that the estimate looses its quality for higher dimensions (note that the right hand side of equation \eqref{eq:estimate_RUO} is proportional to $d^6$).
\begin{figure}[h!]
\begin{subfigure}{1.0\textwidth}
\centering
\subcaption{Convergence of two iterated random unitary operations in qubit systems.}
\if\compileimages1
\begin{tikzpicture}
\begin{semilogyaxis}[
    ylabel={$\| T^n - \mathcal{P} \|$},
    xlabel={iteration},
    xmin=0, xmax=30,
    ymin=1e-10,
    grid=both,
    width=0.8\textwidth,
    height=0.5\textwidth,
    /pgf/number format/.cd,
]
\addplot[color=black, only marks, mark=x,restrict x to domain=0:15]
    table[x=i,y=1]{convergence_2d.txt};
\addplot[color=black,domain=0:15] {4*14*0.08044643^x};
\addplot[color=red, only marks, mark=x]
    table[x=i,y=2]{convergence_2d.txt};
\addplot[color=red,domain=0:50] {4*14*0.45174370^x};
\addlegendentry{$hV,U(V^\dagger)^2,V$};
\addlegendentry{};
\addlegendentry{$(V^\dagger U^\dagger)^3h, UV$};
\addlegendentry{};
\end{semilogyaxis}

\end{tikzpicture}
\else
\includegraphics[width=0.8\textwidth]{convergence_2d}
\fi
\label{fig:convergence_2d}

\end{subfigure}
\begin{subfigure}{1.0\textwidth}
\centering
\subcaption{Convergence of two iterated random unitary operations in qudrit systems.}
\if\compileimages1
\begin{tikzpicture}
\begin{semilogyaxis}[
    ylabel={$\| T^n -\mathcal{P}\|$},
    xlabel={iteration},
    xmin=0, xmax=30,
    ymin=1e-8,
    grid=both,
    width=0.8\textwidth,
    height=0.5\textwidth,
    /pgf/number format/.cd,
]
\addplot[color=black, only marks, mark=x, restrict x to domain=0:55]
    table[x=i,y=1]{convergence_3d.txt};
\addplot[color=black,domain=0:30] {3^2*(3^4-2)*0.55847203^x};
\addplot[color=red, only marks, mark=x]
    table[x=i,y=2]{convergence_3d.txt};
\addplot[color=red,domain=0:30] {3^2*(3^4-2)*0.6740246^x};
\addlegendentry{$h, UV, U$};
\addlegendentry{};
\addlegendentry{$h, UV$};
\addlegendentry{};
\end{semilogyaxis}

\end{tikzpicture}
\else
\includegraphics[width=0.8\textwidth]{convergence_3d}
\fi
\label{fig:convergence_3d}
\end{subfigure}
\caption{Hilbert Schmidt distance between certain iterated random unitary operations and the twirling operation $\mathcal{P}$ from equation \eqref{eq:twirling_op} in dimensions two and three, as a function of the iteration number, plotted in logarithmic scale. The operators $h,U$ and $V$ are defined in section \ref{sec:huv} and were used with parameters from table \ref{tab:params_2d}. The continuous lines represent the estimate given by equation \eqref{eq:estimate_RUO}.}
\end{figure}

\newpage
\section{Summary}
The preparation of Werner states in arbitrary finite dimensional bipartite quantum systems using iterations of random unitary operations has been investigated. Using theory about the asymptotic behaviour of quantum operations, it was shown that the asymptotic eigenvectors of a random unitary operation are determined by the group that is generated by the unitary operators involved in its definition. This group has a natural representation on the underlying Hilbert space. It was shown that if this representation happens to decompose into one dimensional subrepresentations, then each asymptotic eigenvector is diagonal w.r.t. the basis that is defined by the one dimensional subrepresentations.
\\ \\
These results were applied in bipartite qubit systems to obtain a set of random unitary operations that prepare Werner states, which have been compared in terms of convergence rates for the cases of two and three generators. The associated group is isomorphic to SL(2,3), which gives rise to 192 pairs and 1888 triples of elements that generate the whole group and hence yield a random unitary operation that prepares Werner states asymptotically. It was found that 48 of the pairs and 112 of the triples yield a random unitary operation that is not asymptotically stationary.
\\ \\
Then for general $d^2$ dimensional bipartite systems a random unitary operation for the asymptotic preparation of Werner states was constructed, which uses three Kraus operators and depends on six independent parameters that can always be chosen to ensure asymptotic stationarity. The construction gives rise to a whole family of random unitary operations that prepare Werner states in the case of asymptotic stationarity. By multiplying two of the Kraus operators one obtains another random unitary operation that seems to have the same properties, which only uses two Kraus operators. This was proven in the case of the dimension $d$ being odd, the statement possibly generalizes to arbitrary dimensions.
\\ \\
Finally, as an application of the construction in bipartite qubit and qudrit systems, the freedom in six independent parameters was exploited in order to find a random unitary operation with an optimal convergence rate numerically. In the qubit case, the obtained convergence rate was compared to the one obtained from random unitary operations associated with the group SL(2,3). It was found that the special case of the arbitrary dimensional construction yields better values. The resulting random unitary operations have been applied to some example initial state numerically.
\\ \\
The following problems have not been addressed here and need further investigations:
\begin{enumerate}
\item{In the case of two generators of SL(2,3), all resulting random unitary operations were found to be asymptotically stationary if and only if they are diagonalizable, which is hardly a coincidence. In general this is not true, maybe this is due to some special structure in the SL(2,3) case.}
\item{For more than three Kraus operators, the variation of parameters seems to allow for a global minimum of the convergence rate (compare figure \ref{fig:sl23_3d}), which the sequential least squares algorithm that was used in this thesis was not able to attain. Maybe the usage of more refined numerical methods and a closer analysis of the multidimensional surface will result in even better convergence rates.}
\item{How does the usage of more than three Kraus operators affect the convergence rate? This question is related to the remarks that have been pointed out in section \ref{sec:further_ideas}, in particular it is not known so far whether minimizing the number of Kraus operators is actually the most efficient way of asymptotically preparing certain states practically.}
\item{The random unitary operation with two Kraus operators, that was proven to prepare Werner states asymptotically in odd dimension $d$, possibly works in arbitrary dimension as well. Additionally, it may be possible to simplify the proof given in appendix \ref{sec:general_2g_proof}, which relies on a quite lengthy calculation.}
\item{The construction presented in section \ref{sec:huv} involved many arbitrary choices, which are further described in the remarks at the end of the section. A detailed investigation might reveal lots of new independent parameters that can be taken into account in the optimization of the convergence rate. In the cases of qubit and qudrit systems it was found that one parameter has no effect on the convergence rate whatsoever. Maybe this is the case for some new parameters as well, and this can be proven generally.}
\item{As mentioned above, the three generator construction in arbitrary dimension gives rise to a family of random unitary operations for asymptotic preparation of Werner states, namely the ones that match the associated group generated by the unitary Kraus operators. In this work, a particular subset of those has been taken into account for numerical analysis, but this is more to be understood as a proof of concept than a complete discussion. Open questions are: How big is this family exactly? How does it depend on the four independent parameters of the construction? Has the associated group a finite subgroup which suffices for asymptotic preparation of Werner states? Can we say anything about the eigenvalues with modulus strictly smaller than one?}
\item{It is unclear whether in general the associated group of a random unitary operation affects the convergence rate. In the case of SL(2,3) it was found that 96 of 144 pairs of generators yield a random unitary operation with the optimal convergence rate, which is quite a big fraction. This might be a hint to some underlying connection that is yet to be understood.}
\end{enumerate}
\appendix
\section{Appendix: Proof of Theorem \ref{thm:general_2g}} \label{sec:general_2g_proof}
\newtheorem*{Thm4.4}{Theorem 4.4}
\newcommand\req[1]%
  {\stackrel{(#1)}{=}}
\begin{Thm4.4}
Let the dimension $d$ of both partial systems be an odd number, let $\tilde{T}$ be the random unitary operation associated with the Kraus operators $\sqrt{p_1}h$ and $\sqrt{p_2} UV$, with $h, U$ and $V$ from section \ref{sec:huv} and $\alpha \neq 0 \neq \beta$, i.e.
\begin{equation*}
\tilde{T}X = p_1 h_\otimes X h_\otimes^\dagger + p_2 (UV)_\otimes X (UV)_\otimes^\dagger.
\end{equation*}
Then we have $\ker(\tilde{T} - \I) = \mathrm{span}(P_\mathrm{sym},P_\mathrm{asym} )$.
\end{Thm4.4}
\begin{proof}
The action of $(UV)_\otimes$ on the basis \eqref{eq:werner_basis} is given by
\begin{align*}
(UV)_\otimes \ket{\phi_1} &= \mathrm{e}^{2 i \varphi}( \alpha^2 \ket{\phi_2} + \overline{\beta}^2 \ket{\phi_3} - \sqrt{2}\alpha \overline{\beta} \ket{\phi_{2,3}} ) \\ 
(UV)_\otimes \ket{\phi_2} &= \mathrm{e}^{2 i \varphi}( \beta^2 \ket{\phi_2} + \overline{\alpha}^2 \ket{\phi_3} - \sqrt{2}\overline{\alpha} \beta \ket{\phi_{2,3}} ) \\ 
(UV)_\otimes \ket{\phi_k} &= \ket{\phi_{k+1}} \\
(UV)_\otimes \ket{\phi_{1,2}} &= \mathrm{e}^{2 i \varphi}( \sqrt{2} \alpha \beta \ket{\phi_2} - \sqrt{2} \overline{\alpha} \overline{\beta} \ket{\phi_3} + (|\alpha|^2 - |\beta|^2) \ket{\phi_{2,3}}) \\
(UV)_\otimes \ket{\psi_{1,2}} &= \mathrm{e}^{2 i \varphi} \ket{\psi_{2,3}} \\
(UV)_\otimes \ket{\phi_{1,k}} &= \mathrm{e}^{i \varphi} (\alpha \ket{\phi_{2,k+1}} - \overline{\beta}\ket{\phi_{3,k+1}}) \\
(UV)_\otimes \ket{\phi_{2,k}} &= \mathrm{e}^{i \varphi} (\beta \ket{\phi_{2,k+1}} + \overline{\alpha}\ket{\phi_{3,k+1}}) \\
(UV)_\otimes \ket{\phi_{k,l}} &= \ket{\phi_{k+1,l+1}} \text{ if }2<k<l,
\end{align*}
where $k>2$ and the last three equations hold analogously for $\ket{\psi_{i,j}}$ instead of $\ket{\phi_{i,j}}$. Here the index $k=d+1$ is identified with $k=1$. Let $X \in \ker(\tilde{T}-\I)$. As in the proof of Theorem \ref{thm:general_3g}, Theorem \ref{thm:quantumop_eigenspace} yields that $X$ is a simultaneous eigenvector of $h_\otimes.h_\otimes^\dagger$ and $(UV)_\otimes . (UV)_\otimes^\dagger$. Since $h_\otimes$ has finite order, the Theorems \ref{thm:eigenvector_diagonal} and \ref{thm:eigenspace_group} imply that the only non-vanishing matrix elements of $X$ in the basis \eqref{eq:werner_basis} are the diagonal ones as well as $\bra{\phi_{i,j}} X \ket{\psi_{i,j}}$ and $\bra{\psi_{i,j}} X \ket{\phi_{i,j}}$ for all $i < j$. 
\\ \\
We focus on diagonality first. For convenience of notation define $p \coloneqq |\alpha|^2$, $q \coloneqq |\beta|^2$ and $\braket{i,j} \coloneqq \bra{\phi_{i,j}} X \ket{\psi_{i,j}}$. Note that we have $p+q = 1$, therefore the equations
\begin{equation}\label{eq:A_pq}
p^2 - q^2 = 1-2q = 2p - 1 = p-q \tag{$pq$}
\end{equation}
are fulfilled. By applying $(UV)_\otimes^\dagger . (UV)_\otimes$ we obtain the equations
\begin{equation}\label{eq:A_1}
\braket{1,k} = p \braket{2,k+1} + q \braket{3,k+1} \tag{1}
\end{equation}
\begin{equation}\label{eq:A_2}
\braket{2,k} = q \braket{2,k+1} + p \braket{3,k+1} \tag{2}
\end{equation}
\begin{equation}\label{eq:A_circ}
\braket{k,l} = \braket{k+1,l+1} \tag{$\circlearrowleft$},
\end{equation}
where $2<k<l$, and
\begin{equation}\label{eq:A_3}
\braket{1,2} = (q-p) \braket{2,3}. \tag{3}
\end{equation}
We claim that the equations \eqref{eq:A_1}, \eqref{eq:A_2} and \eqref{eq:A_circ} imply $\braket{k,l} = \braket{1,2}$ for all $k<l$. To begin with, note that the equations \eqref{eq:A_1} and \eqref{eq:A_2} immediately imply
\begin{equation}\label{eq:A_d}
\braket{1,d} = p \braket{1,2} + q \braket{1,3} \text{ and } \braket{2,d} = q \braket{1,2} + p \braket{1,3}. \tag{$d$}
\end{equation}
Then we find for any fixed $k>2$ by \eqref{eq:A_circ}
\begin{equation}\label{eq:A_4}
\braket{3,k+1} = \braket{3+d-(k+1),d} = \braket{1,d-k+3},\tag{4}
\end{equation}
whence substituting \eqref{eq:A_2} in \eqref{eq:A_1} gives
\begin{align}\label{eq:A_5}
\begin{split}
\braket{1,k} &= p\left(\frac{1}{q} \braket{2,k} - p\braket{3,k+1} \right) + q \braket{3,k+1} = \frac{p}{q}\braket{2,k} + \left(q - \frac{p^2}{q}\right)\braket{3,k+1} \\
&\req{pq),(4} \frac{p}{q} \braket{2,k} + \left(1-\frac{p}{q}\right)\braket{1,d-k+3}.
\end{split}\tag{5}
\end{align}
Further we find by applying \eqref{eq:A_1} and \eqref{eq:A_2} successively
\begin{equation}\label{eq:A_sum1}
\braket{1,k} = p q^{d-k-1} \braket{2,d} + q \braket{1,d-k+3} + \frac{p^2}{q} \sum \limits_{i=1}^{d-k-1} q^i \braket{1,d-k+3-i} \tag{$\Sigma1$}
\end{equation}
and analogously
\begin{equation}\label{eq:A_sum2}
\braket{2,k} = q^{d-k} \braket{2,d} + p \braket{1,d-k+3} + p \sum \limits_{i=1}^{d-k-1} q^i \braket{1,d-k+3-i}. \tag{$\Sigma2$}
\end{equation}
This can of course be shown rigorously using induction. Note that at this step we have reduced the problem of identifying all matrix elements $\braket{k,l}$ to just identifying the matrix elements $\braket{1,l}$. Indeed, if $\braket{1,l} = \braket{1,2}$ for all $l>2$ then we have $\braket{2,d} = \braket{1,2}$ by equation \eqref{eq:A_d}, and using a geometric sum equation \eqref{eq:A_sum2} simplifies to
\begin{align*}
\braket{2,k} &= \left( q^{d-k} + p + p \sum \limits_{i=1}^{d-k-1} q^i \right) \braket{1,2} \\
&= \left( q^{d-k} + p + p\left( \frac{1-q^{d-k}}{1-q}-1\right) \right) \braket{1,2} = \braket{1,2} \text{ for all }k>2.
\end{align*}
All other matrix elements are taken care of by equation \eqref{eq:A_circ}. It therefore suffices to show that all matrix elements $\braket{1,l}$ coincide. In order to achieve that, we start with investigating the further consequences of equation \eqref{eq:A_sum1}:
\begin{align}\label{eq:A_6}
\begin{split}
q\braket{1,k+1} - \braket{1,k} &\req{\Sigma1} q \left(q\braket{1,d-k+2} + \frac{p^2}{q} \sum \limits_{i=1}^{d-k-2} q^i \braket{1,d-k+2-i} \right) \\ 
&\hspace{0.6cm}- \left( q \braket{1,d-k+2} + \frac{p^2}{q} \sum \limits_{i=1}^{d-k-1} q^i \braket{1,d-k+3-i} \right) \\
&= q^2 \braket{1,d-k+2} - q \braket{1,d-k+3} - p^2 \braket{1,d-k+2} \\
&\req{pq} (1-2p)\braket{1,d-k+2} - q\braket{1,d-k+3}. 
\end{split}\tag{6}
\end{align}
Now we apply equation \eqref{eq:A_6} on itself for $d-k >0$ and find
\begin{align*}
q\braket{1,k+1} - \braket{1,k} &\req{6} (1-2p)\braket{1,d-k+2} - q\braket{1,d-k+3} \\
&= - (q\braket{1,d-k+3} - \braket{1,d-k+2}) - 2p \braket{1,d-k+2} \\
&\req{6} -((1-2p)\braket{1,k} - q\braket{1,k+1}) - 2p \braket{1,d-k+2} \\
&= q\braket{1,k+1} - \braket{1,k} + 2p( \braket{1,k} - \braket{1,d-k+2} ).
\end{align*}
Solving for $\braket{1,k}$ yields
\begin{equation}\label{eq:A_7}
\braket{1,k} = \braket{1,d-k+2} \text{ for all }2<k<d.\tag{7}
\end{equation}
This strengthens equation \eqref{eq:A_4}, which then reads
\begin{equation}\label{eq_A_8}
\braket{3,k+1} \req{4} \braket{1,d-k+3} \req{7} \braket{1,k-1} \text{ for all }3<k<d.\tag{8}
\end{equation}
Since we assumed $d$ to be odd, by equation \eqref{eq:A_7} we find $\braket{1,j-1}=\braket{1,j}$ for $j \coloneqq \frac{d+3}{2}$ (consider for instance $d=7$, then we have $\braket{1,3} = \braket{1,6}$ and $\braket{1,4} = \braket{1,5}$, so $j=5=\frac{7+3}{2}$.) This implies $\braket{2,j} = \braket{2,j+1}$ since it holds
\begin{equation*}
\braket{1,k}+\braket{2,k} \req{1),(2} \braket{2,k+1} + \braket{3,k+1} \req{8} \braket{2,k+1} + \braket{1,k-1}.
\end{equation*}
We now show by induction that $\braket{1,j-k} = \braket{2,j}$ holds for all $1\leq k<j-3$. For $k=1$ we find
\begin{equation*}
p\braket{3,j+1} \req{2} \braket{2,j}-q\braket{2,j+1} = (1-q) \braket{2,j},
\end{equation*}
thus it follows
\begin{equation*}
\braket{2,j} = \braket{3,j+1} \req{8} \braket{1,j-1}.
\end{equation*}
Let $2\leq k<j-3$ and assume $\braket{1,j-l} = \braket{2,j}$ for all $1 < l \leq k$. We find, applying equation \eqref{eq:A_2} successively, as in the derivation of equation \eqref{eq:A_sum1},
\begin{align*}
\braket{2,j} &= \braket{1,j-k} \req{1} p \braket{2,j-k+1} + q \braket{3,j-k+1} \\
&\req{2),(8} pq^{k-1} \braket{2,j} + q \braket{1,j-(k+1)} + \frac{p^2}{q} \sum \limits_{i=1}^{k-1} q^i \underbrace{\braket{3,j-k+1+l}}_{\req{8} \braket{1,j-(k+1-l)} = \braket{2,j}}, \\
\end{align*}
which yields
\begin{align*}
\braket{1,j-(k+1)} &= \frac{1}{q} \left( 1-pq^{k-1} - \frac{p^2}{q} \sum \limits_{i=1}^{k-1} q^i \right) \braket{2,j} \\
&= \frac{1}{q} \left( 1-pq^{k-1} - \frac{p^2}{q} \left( \frac{1-q^k}{1-q} - 1 \right) \right) \braket{2,j} = \braket{2,j}.
\end{align*}
Now we are basically done. As we have
\begin{equation*}
\braket{2,j} = \braket{1,3} \req{8} \braket{1,d-1} \req{1} p \braket{2,d} + q \underbrace{\braket{3,d}}_{=\braket{1,4} = \braket{2,j}}
\end{equation*}
it follows $\braket{2,j} = \braket{2,d}$ and thus by equation \eqref{eq:A_d} we conclude $\braket{1,d} = \braket{2,j} = \braket{1,2}$. We now have proven the claim that the equations \eqref{eq:A_1}, \eqref{eq:A_2} and \eqref{eq:A_circ} imply that all matrix elements $\braket{k,l}$ coincide. To conclude that $X$ is diagonal, observe that equation \eqref{eq:A_3} fixes this value to be 0, as $p - q \neq 1$.
\\ \\
It remains to show that the diagonal matrix elements corresponding to $P_\mathrm{sym}$ as well as those corresponding to $P_\mathrm{asym}$ coincide. We are actually already done with the case of $P_\mathrm{asym}$, as denoting $\bra{\psi_{k,l}}X\ket{\psi_{k,l}} = \braket{k,l}$ gives exactly the equations \eqref{eq:A_1}, \eqref{eq:A_2} and \eqref{eq:A_circ} once again. The same thing works for the other diagonal matrix elements $\bra{\phi_{k,l}}X\ket{\phi_{k,l}} \eqqcolon \braket{k,l}$, though we also have the matrix elements $\bra{\phi_k}X\ket{\phi_k} \eqqcolon \braket{k}$ to take into account. In addition to the equations \eqref{eq:A_1}, \eqref{eq:A_2} and \eqref{eq:A_circ} we have
\begin{equation}\label{eq:A_9}
\braket{1} = p^2 \braket{2} + q^2 \braket{3} + 2 pq \braket{2,3},\tag{9}
\end{equation}
\begin{equation}\label{eq:A_10}
\braket{2} = q^2 \braket{2} + p^2 \braket{3} + 2 pq \braket{2,3},\tag{10}
\end{equation}
\begin{equation}\label{eq:A_11}
\braket{k} = \braket{k+1} \text{ for all }k>2\tag{11}
\end{equation}
and, replacing equation \eqref{eq:A_3},
\begin{equation}\label{eq:A_12}
\braket{1,2} = 2pq( \braket{2} + \braket{3}) + (p-q)^2 \braket{2,3}.\tag{12}
\end{equation}
Since $\braket{2,3} = \braket{1,2}$ and $\braket{3} = \braket{1}$ it follows
\begin{equation}\label{eq:A_13}
\braket{1,2} = \frac{2pq}{1-(p-q)^2}(\braket{1} + \braket{2}) \req{pq} \frac{1}{2}(\braket{1} + \braket{2}).\tag{13}
\end{equation}
This finally yields
\begin{equation*}
\braket{2} \req{10),(13} q^2 \braket{2} + p^2 \braket{1} + pq (\braket{1} + \braket{2}) = p \braket{1} + q \braket{2} = \braket{1}
\end{equation*}
and taking care of all other matrix elements is trivial, therefore the proof is complete.
\end{proof}
\newpage
\section{Appendix: Numerical evaluation data}\label{ch:numerical_appendix}
\begin{figure}[h!]
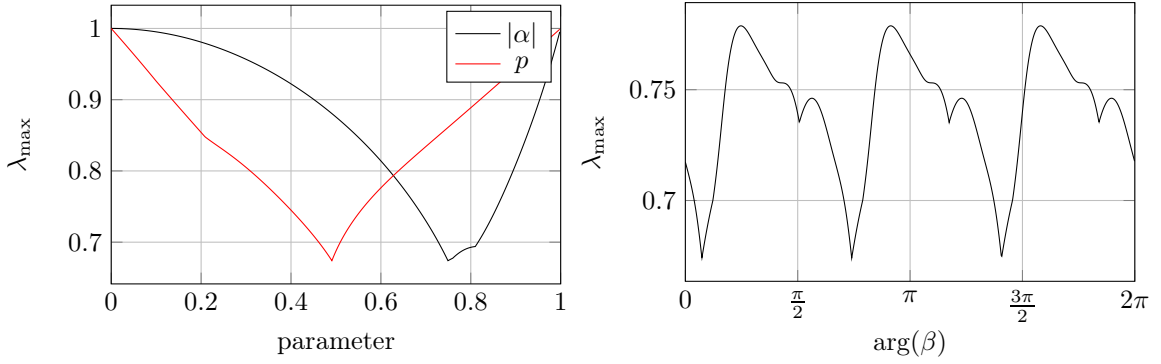
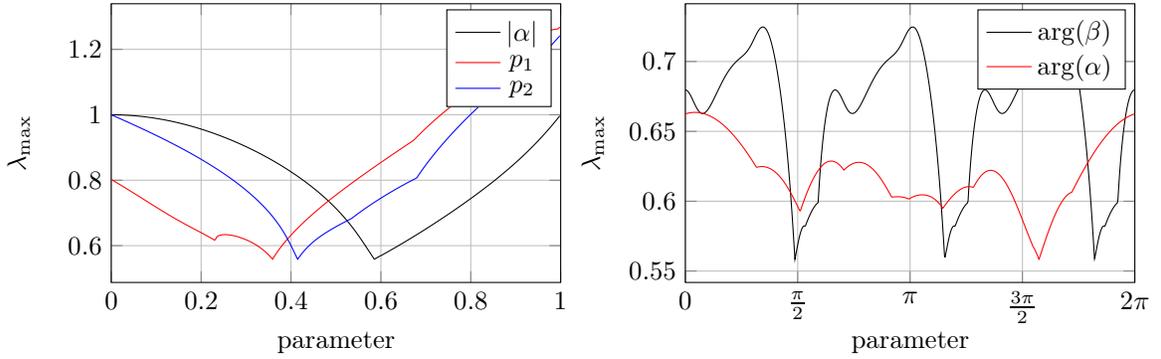

\centering
\begin{subfigure}{1.0\textwidth}
\if\compileimages1
\begin{minipage}{0.5\textwidth}
\begin{tikzpicture}
\begin{axis}[
    xlabel={parameter},
    ylabel={$\lambda_\mathrm{max}$},
    xmin=0, xmax=1,
    grid=both,
    width=\textwidth,
    height=0.7\textwidth,
    /pgf/number format/.cd,
]
 
\addplot[color=black]
    table[x=x,y=|alpha|]{maxeig_d3g2_prange.txt};
\addlegendentry{$|\alpha|$}
\addplot[color=red]
    table[x=x,y=p1]{maxeig_d3g2_prange.txt};
\addlegendentry{$p$}
\end{axis}

\end{tikzpicture}
\end{minipage}
\begin{minipage}{0.5\textwidth}
\begin{tikzpicture}
\begin{axis}[
    xlabel={$\arg(\beta)$},
    ylabel={$\lambda_\mathrm{max}$},
    xmin=0, xmax=2*pi,
    xtick={0, 1.570796, 3.141593, 4.712389, 6.28319},
    xticklabels={0, $\frac{\pi}{2}$, $\pi$, $\frac{3\pi}{2}$, $2\pi$},
    grid=both,
    width=\textwidth,
    height=0.7\textwidth,
    /pgf/number format/.cd,
]
 
\addplot[color=black]
    table[x=x,y=arg(beta)]{maxeig_d3g2_phirange.txt};
\end{axis}
\end{tikzpicture}
\end{minipage}
\else
\includegraphics[width=0.5\textwidth]{maxeig_d3g2_prange}
\includegraphics[width=0.5\textwidth]{maxeig_d3g2_phirange}
\fi
\subcaption{The convergence parameter $\lambda_\mathrm{max}$ of the random unitary operation associated with the generators $h$ and $UV$ in dimension $d = 3$. A variation of the missing parameters $\varphi$ and $\arg(\alpha)$ leaves $\lambda_\mathrm{max}$ unchanged, which are therefore omitted.}
\label{fig:maxeig_d3g2}
\vspace{1cm}
\end{subfigure}
\begin{subfigure}{1.0\textwidth}
\if\compileimages1
\begin{minipage}{0.5\textwidth}
\begin{tikzpicture}
\begin{axis}[
    xlabel={parameter},
    ylabel={$\lambda_\mathrm{max}$},
    xmin=0, xmax=1,
    grid=both,
    width=\textwidth,
    height=0.7\textwidth,
    /pgf/number format/.cd,
]
 
\addplot[color=black]
    table[x=x,y=|alpha|]{maxeig_d3g3_prange.txt};
\addlegendentry{$|\alpha|$}
\addplot[color=red]
    table[x=x,y=p1]{maxeig_d3g3_prange.txt};
\addlegendentry{$p_1$}
\addplot[color=blue]
    table[x=x,y=p2]{maxeig_d3g3_prange.txt};
\addlegendentry{$p_2$}
\end{axis}
\end{tikzpicture}
\end{minipage}
\begin{minipage}{0.5\textwidth}
\begin{tikzpicture}
\begin{axis}[
    xlabel={parameter},
    ylabel={$\lambda_\mathrm{max}$},
    xmin=0, xmax=2*pi,
    xtick={0, 1.570796, 3.141593, 4.712389, 6.28319},
    xticklabels={0, $\frac{\pi}{2}$, $\pi$, $\frac{3\pi}{2}$, $2\pi$},
    grid=both,
    width=\textwidth,
    height=0.7\textwidth,
    /pgf/number format/.cd,
]
 
\addplot[color=black]
    table[x=x,y=arg(beta)]{maxeig_d3g3_phirange.txt};
\addlegendentry{$\arg(\beta)$}
\addplot[color=red]
    table[x=x,y=arg(alpha)]{maxeig_d3g3_phirange.txt};
\addlegendentry{$\arg(\alpha)$}
\end{axis}
\end{tikzpicture}
\end{minipage}
\else
\includegraphics[width=0.5\textwidth]{maxeig_d3g3_prange}
\includegraphics[width=0.5\textwidth]{maxeig_d3g3_phirange}
\fi
\subcaption{The convergence parameter $\lambda_\mathrm{max}$ of the random unitary operation associated with the generators $h, UV$ and $U$ in dimension $d=3$. Once again, variation of $\varphi$ has no effect on $\lambda_\mathrm{max}$.}
\label{fig:maxeig_d3g3}
\end{subfigure}
\caption{The convergence parameter $\lambda_\mathrm{max}$ of the random unitary operation associated with different tuples of generators in a neighbourhood of the optimization minimum as a one dimensional function of each individual parameter, in dimension $d=3$. The parameters can be found in table \ref{tab:params_2d}.}
\end{figure}
\newpage
\begin{table}[!htb]
\begin{center}
 \caption{Parameters of the random unitary operations resulting in the best value for $\lambda_\mathrm{max}$ in dimensions two and three.}
 \label{tab:params_top10}
 \scalebox{0.8}{
 \begin{tabular}{c c c c c c c c c}
 \hline \hline
  $d$ & generators & $\lambda_\mathrm{max}$ & $\varphi$ & $|\alpha|$ & $\arg( \alpha )$ & $\arg( \beta )$ & $p_1$ & $p_2$ \\ \hline
2 & $(V^\dagger U^\dagger)^3 h,UV$ & 0.45171364(1) & 0.02040734 & 0.93254821 & 2.89447138 & 4.87953210 & 0.48788387 & - \\
2 & $V^\dagger U^\dagger h(V^\dagger U^\dagger)^2 ,UV$ & 0.45177443(1) & 2.22634490 & 0.93273780 & 0.56307152 & 1.73613282 & 0.48924909 & - \\
2 & $(V^\dagger U^\dagger)^2 hV^\dagger U^\dagger ,UV$ & 0.45182053(1) & 2.43540064 & 0.93236651 & 1.66719716 & 1.73922016 & 0.48660551 & - \\
2 & $h(V^\dagger U^\dagger)^3,UV$ & 0.45247260(1) & 4.66621475 & 0.93266531 & 5.80165408 & 1.75004489 & 0.48762139 & - \\
2 & $(h^\dagger)^2 UVh,h$ & 0.47131080(1) & 5.51693231 & 0.93397081 & 6.03814570 & 4.93978785 & 0.54533411 & - \\
2 & $h,UV$ & 0.47132816(1) & 4.41035379 & 0.93329073 & 0.01310998 & 4.13663214 & 0.45454826 & - \\
2 & $h^2 UV,h$ & 0.47142342(1) & 2.02524109 & 0.93459679 & 1.45866385 & 2.59787038 & 0.54513095 & - \\
2 & $UVhV^\dagger U^\dagger ,UV$ & 0.47156322(1) & 5.19279314 & 0.93309190 & 1.74555195 & 4.13291884 & 0.45454065 & - \\
2 & $V^\dagger U^\dagger hUV,UV$ & 0.47186921(1) & 4.47128947 & 0.93289846 & 0.14492048 & 4.13217062 & 0.45463401 & - \\
2 & $UVh^\dagger ,h$ & 0.47214326(1) & 4.49293324 & 0.93363022 & 2.91434885 & 1.80069191 & 0.54492723 & - \\
2 & $hV,U (V^\dagger)^2,V$ & 0.08044643(1) & 4.18274534 & 0.91986392 & 1.00676113 & 2.40909110 & 0.22305894 & 0.42044878 \\
2 & $Vh,(V^\dagger)^2 U,V$ & 0.08613587(1) & 5.29667689 & 0.92036449 & 4.14864868 & 0.73242149 & 0.22279991 & 0.42048974 \\
2 & $VhV^\dagger ,VU,V$ & 0.08952903(1) & 6.20626095 & 0.41828259 & 0.76691346 & 6.15870903 & 0.42214212 & 0.13221572 \\
2 & $Vh,V^\dagger U,V$ & 0.09185870(1) & 5.85279091 & 0.37513429 & 1.39285807 & 5.45805530 & 0.43012512 & 0.44798363 \\
2 & $V^2 h,UV^\dagger ,V$ & 0.09211565(1) & 4.02890429 & 0.78175738 & 4.83014038 & 3.92323397 & 0.38164265 & 0.34479035 \\
2 & $hV,VU,V$ & 0.09442135(1) & 0.32144573 & 0.89702868 & 0.85925184 & 5.64251529 & 0.13668014 & 0.44250114 \\
2 & $hV,UV,V$ & 0.09645365(1) & 3.45912305 & 0.90414785 & 4.35828799 & 0.78092410 & 0.31909686 & 0.41524112 \\
2 & $h,(V^\dagger)^2 UV^\dagger ,V$ & 0.10979778(1) & 2.66859740 & 0.42807350 & 0.75082334 & 3.06736006 & 0.41127768 & 0.16881174 \\
2 & $Vh,UV,V$ & 0.10979894(1) & 0.79450488 & 0.86364152 & 2.49111116 & 2.48950978 & 0.14233361 & 0.44102029 \\
2 & $V^\dagger h,UV^\dagger ,V$ & 0.11122951(1) & 1.75306750 & 0.56171650 & 5.23958689 & 4.91760721 & 0.33595959 & 0.30103239 \\
3 & $h,UV$ & 0.67402461(1) & 2.86002806 & 0.74921865 & 3.66908666 & 2.32545709 & 0.49097422 & - \\
3 & $(h^\dagger)^2 UV,h$ & 0.67402857(1) & 3.58040315 & 0.74919715 & 0.63798609 & 4.42006854 & 0.50905687 & - \\
3 & $hUV,h$ & 0.67410727(1) & 1.30053265 & 0.74916229 & 6.15894473 & 1.27787623 & 0.50896175 & - \\
3 & $UV(h^\dagger)^2,h$ & 0.67416358(1) & 2.61522285 & 0.74952701 & 6.0850070 & 2.32701916 & 0.50938744 & - \\
3 & $h^\dagger UVh^\dagger ,h$ & 0.67418091(1) & 5.02413306 & 0.74901570 & 2.92768346 & 2.32504626 & 0.50911867 & - \\
3 & $h^2 UV,h$ & 0.67427969(1) & 1.98220559 & 0.74996984 & 2.90787258 & 0.22978137 & 0.50864663 & - \\
3 & $UVhV^\dagger U^\dagger ,UV$ & 0.67438372(1) & 0.12493273 & 0.75030231 & 3.56111816 & 4.41811004 & 0.49146879 & - \\
3 & $UVh^\dagger ,h$ & 0.67440881(1) & 1.07883319 & 0.74905991 & 2.41860741 & 5.47437286 & 0.50996725 & - \\
3 & $UVh,h$ & 0.67495927(1) & 1.53261092 & 0.74854204 & 0.45067027 & 1.29627404 & 0.51126798 & - \\
3 & $h^\dagger UV,h$ & 0.67541822(1) & 5.78498571 & 0.75226065 & 1.53919481 & 1.26622318 & 0.50428252 & - \\
3 & $h,UV,U$ & 0.55847203(1) & 0.36470858 & 0.58546063 & 4.94276220 & 1.53168497 & 0.35903090 & 0.41473933 \\
3 & $h,(U^\dagger)^2 V,U$ & 0.56003566(1) & 4.58127581 & 0.57185926 & 2.67573573 & 5.67626474 & 0.32911909 & 0.39704823 \\
3 & $h,VU,U$ & 0.56253792(1) & 2.46911311 & 0.52749982 & 5.50102328 & 5.99468805 & 0.34919650 & 0.41712141 \\
3 & $V^2 h,UV,V$ & 0.56455353(1) & 2.57411913 & 0.47752815 & 2.06239933 & 3.46019409 & 0.30865684 & 0.40848157 \\
3 & $h,V^\dagger UV^2,V$ & 0.56553419(1) & 4.58450643 & 0.59397997 & 3.94475491 & 4.90128165 & 0.39833521 & 0.39098377 \\
3 & $h,V(U^\dagger)^2 ,U$ & 0.56574477(1) & 2.23336623 & 0.63624897 & 5.11453739 & 5.75914870 & 0.29210269 & 0.43785722 \\
3 & $h,(V^\dagger)^2 U,V$ & 0.56615350(1) & 2.94321676 & 0.30747956 & 6.22028768 & 5.94651517 & 0.42198975 & 0.41244825 \\
3 & $h,UV^2,V$ & 0.56742223(1) & 3.29460007 & 0.44631186 & 5.73793993 & 3.68805308 & 0.38705269 & 0.35993853 \\
3 & $h,VUV,V$ & 0.56927646(1) & 1.50364426 & 0.42866536 & 2.92438845 & 1.65786749 & 0.39764834 & 0.39608258 \\
3 & $h,VU(V^\dagger)^2 ,V$ & 0.57125921(1) & 4.05121612 & 0.54547945 & 5.44226723 & 3.71599001 & 0.30612825 & 0.43194246 \\
  \hline \hline
 \end{tabular}}
\end{center}
\end{table}
\newpage
\bibliographystyle{unsrt}
\bibliography{Stonner_Construction-of-RUO-for-Asymptotic-Preparation-of-Werner-States}


\end{document}